\documentclass[aps,prx,reprint,superscriptaddress,
floatfix]{revtex4-2}

\usepackage[T1]{fontenc}
\usepackage{mathptmx}

\usepackage{amsmath,amsfonts,amssymb,upgreek}
\usepackage{braket}

\usepackage{graphicx}
\usepackage{booktabs}
\usepackage{multirow}

\usepackage{hyperref}
\hypersetup{
  colorlinks=true,
  linkcolor=[rgb]{0.19,0.52,0.67},
  urlcolor=[rgb]{0.12,0.29,0.16},
  citecolor=[rgb]{0.75,0.37,0.14}
}

\newcommand{\ee}{\mathrm{e}}
\newcommand{\ii}{\mathrm{i}}

\begin{document}
\title{Distributed Phase Sensing with Multiphoton States in Optical Interferometry}
\author{Subhrajit Modak}
\thanks{These authors contributed equally to this work}
\affiliation{Dipartimento Interateneo di Fisica, Politecnico di Bari, 70126 Bari, Italy}

\author{Danilo Triggiani}
\affiliation{Dipartimento Interateneo di Fisica, Politecnico di Bari, 70126 Bari, Italy}
\affiliation{Istituto Nazionale di Fisica Nucleare (INFN), Sezione di Bari, 70126 Bari, Italy}

\author{Cosmo Lupo}
\affiliation{Dipartimento Interateneo di Fisica, Politecnico di Bari, 70126 Bari, Italy}
\affiliation{Istituto Nazionale di Fisica Nucleare (INFN), Sezione di Bari, 70126 Bari, Italy}

\begin{abstract}
We investigate interferometric phase-estimation using separable photon inputs that evolve into number-path entangled states through linear optical networks, followed by photon-number-resolving detection. A simple analytical expression for the classical Fisher information at zero phase is derived for arbitrary $N$-photon states distributed across 2$N$ optical modes, partitioned into phase-encoding and reference blocks. Among all possible photon distributions between these blocks, the balanced configuration maximizes the phase sensitivity for every photon number $N$ and uniquely exhibits a phase-independent response. The achievable sensitivity degrades monotonically with increasing asymmetry in the photon distribution. We further investigate the robustness of the protocol in the presence of realistic photon loss and extend the analysis to distributed architectures with multiple receivers. In the low photon-flux regime, vacuum fluctuations fundamentally limit local quadrature measurements, whereas nonlocal photon-number-resolving measurements exploit multiphoton interference to mitigate loss-induced sensitivity degradation. Together, these results establish a scalable framework for quantum-enhanced distributed multimode metrology.

\end{abstract}

\maketitle
\section{Introduction}\label{sec:intro}

Quantum interferometry provides a cornerstone for precision metrology, where the achievable phase sensitivity is ultimately constrained by the interplay between quantum coherence, measurement strategy, and noise~\cite{helstrom1969quantum,PhysRevLett.96.010401}. Since the early recognition that non-classical states can surpass classical shot-noise scaling~\cite{PhysRevA.33.4033,PhysRevD.23.1693}, a wide range of photonic implementations has been developed to exploit quantum resources for enhanced sensing and distributed phase estimation~\cite{giovannetti2004quantum,Dowling01032008,nagata2007beating,PhysRevLett.102.100401}. In particular, single-photon interferometric protocols have emerged as minimal-resource platforms for probing the fundamental role of path superposition and measurement-induced correlations in quantum-enhanced metrology~\cite{RevModPhys.79.135,nielsen2010quantum}.

Early works established that entangled resources can surpass classical precision limits~\cite{PhysRevLett.85.2733,doi:10.1126/science.1138007}, with paradigmatic examples including NOON states and entangled probes in Mach-Zehnder interferometers~\cite{PhysRevA.33.4033,PhysRevLett.71.1355,KIM199837,RevModPhys.90.035005}. Despite their favorable metrological properties, the preparation and manipulation of large-photon-number entangled states remain experimentally challenging~\cite{dowling2008quantum,walther2004broglie}, motivating the search for scalable alternatives based on experimentally accessible photonic resources. A promising direction is provided by multimode linear-optical networks, where single photons are coherently distributed across many optical paths~\cite{tillmann2013experimental}. Such transformations generate number-path entanglement using only passive optical elements, eliminating the need for nonlinear interactions or complex multiphoton state preparation~\cite{aaronson2011computational,tillmann2013experimental}. More recently, it has become clear that the metrological power of these architectures depends not only on the quantum state itself but also on the measurement strategy used to extract phase information~\cite{demkowicz2015quantum,PhysRevLett.106.153603}.

Here, we investigate a broad class of multimode interferometric phase-estimation protocols employing single-photon inputs, linear-optical evolution, and photon-number-resolving detection~\cite{PhysRevA.80.043822}. The photons are partitioned into phase-encoding and reference sectors, and the resulting interference is analyzed through a general linear-optical measurement network. We express the output probabilities in terms of matrix permanents~\cite{10.1098/rspa.2011.0232}, highlighting the role of many-photon interference in determining the metrological sensitivity of the protocol. We show that the attainable phase sensitivity is governed by the partition of photons between the phase and reference sectors. This result follows from an analytical expression for the classical Fisher information (CFI) at zero phase, derived for arbitrary $N$-photon input states. Among all configurations, the balanced partition achieves the highest sensitivity for any photon number $N$ and is uniquely characterized by a phase-independent response. In contrast, increasing asymmetry in the photon distribution leads to a monotonic decrease in sensitivity. We further investigate the robustness of these protocols against photon loss in the transmission channel, a key limitation for practical quantum-enhanced metrology~\cite{demkowicz2012elusive,demkowicz2015quantum,escher2011general}. The achievable precision remains strongly dependent on the measurement strategy. In the low-photon-flux regime, local Gaussian measurements, such as independent homodyne detection, are constrained by vacuum fluctuations, resulting in reduced sensitivity. However, nonlocal photon-number-resolving measurements directly exploit multiphoton interference and retain a larger fraction of the available phase information, providing enhanced robustness against photon loss.

\begin{figure*}[t]
\centering
\includegraphics[width=.73\linewidth,height=2.9in]{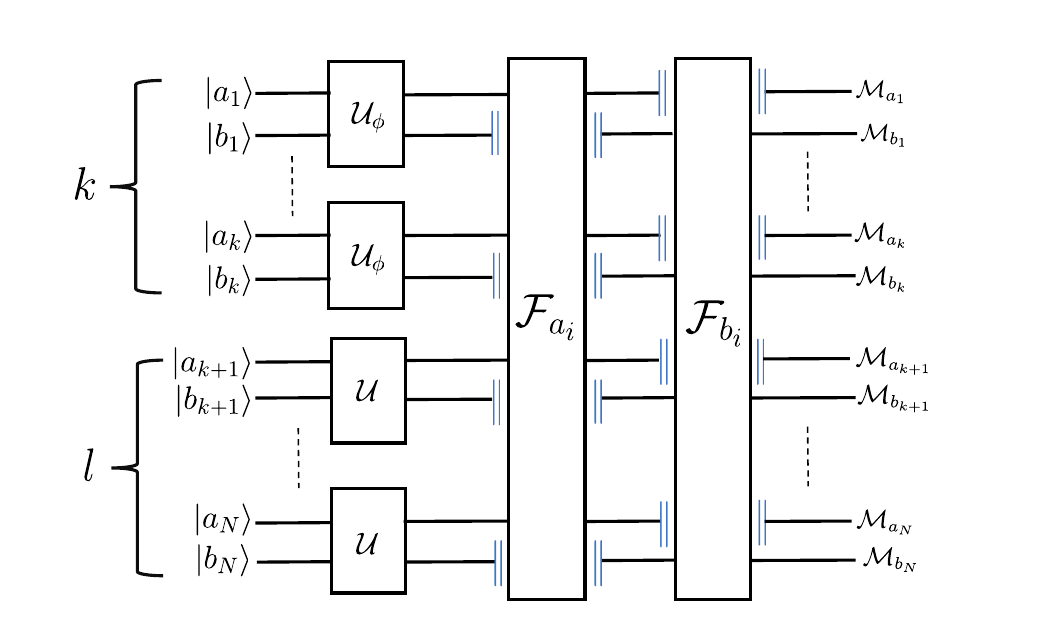}
\caption{Schematic representation of a linear-optical interferometric architecture for multiphoton phase estimation. The input state consists of $N$ single photons distributed across $2N$ optical modes. The photons evolve under a global unitary transformation comprises a phase-dependent interferometric network that simultaneously generates multipartite path entanglement and encodes the unknown phase parameter $\phi$, followed by a quantum Fourier transform $\mathcal{F}$, which coherently redistributes the optical amplitudes among all modes. Within the phase-encoding network, individual interferometric blocks implement either the phase shift $\phi$ or the identity operation, thereby imprinting the phase information onto the entangled quantum state. At the output, photon-number-resolving detection measure all possible occupation-number configurations, yielding a probability distribution from which the phase $\phi$ can be inferred. The schematic illustrates the ideal unitary evolution, however, practical implementations are subject to transmission loss that limit the achievable precision.}

\label{vp}
\end{figure*}
Finally, we extend the analysis to distributed sensing architectures in which a single photon is coherently shared among multiple spatially separated receivers~\cite{hong2021quantum}. Such configurations naturally realize telescope-like architectures, where distributed optical modes acquire phase information from an incoming wavefront and are subsequently recombined through a global interferometric network~\cite{Huang31122026,PhysRevA.108.052408,PhysRevLett.131.210801,354q-ch63}. By accessing multiple baselines simultaneously, these architectures enable enhanced estimation of spatially encoded phases and highlight the role of resource distribution and measurement nonlocality in multimode quantum-enhanced metrology.

The paper is organized as follows. Section~\ref{inb} introduces the linear-optical interferometric architecture and derives a general expression for the CFI at the zero-phase point as a function of the partition between the phase-encoding and reference blocks. Section~\ref{inc} examines the effect of photon loss and shows that, at zero phase, the CFI is suppressed by the probability of retaining all photons, whereas away from this point, the loss dependence becomes more intricate. Section~\ref{ind} compares these results with local homodyne detection, highlighting the role of measurement nonlocality. Section~\ref{ine} extends the framework to multimode distributed sensing with a single photon shared among multiple receivers. Finally, Sec.~\ref{ss} summarizes the main results and provides an outlook on future directions.

\section{Interferometric Scheme}\label{inb}

We discuss a general scheme of linear-optical interferometric architecture for phase estimation, where an $N$-photon input state is distributed over $2N$ modes arranged into $N$ pairs $\{a_i,b_i\}$, as shown in \figurename~\ref{vp}. 
Each pair is processed by an independent two-mode unitary that generates path entanglement while possibly encoding an unknown phase $\phi$~\cite{PhysRevLett.100.073601,PhysRevA.109.053508}
\begin{equation}
	\ket{\psi} =  \frac{1}{2^{N/2}}\prod_{i=1}^{k}(\hat{a}_i^\dag+\ee^{\ii\phi}\hat{b}_i^\dag)\prod_{j=k+1}^{N}(\hat{a}_j^\dag+\hat{b}_j^\dag)\ket{0}.
\end{equation}
The number of phase-encoding blocks, $k$, treated as a tunable resource for optimizing phase-estimation precision, with the $N = k + l$ mode pairs partitioned into $k$ pairs carrying the unknown phase and $l$ phase-independent references. Subsequently, the modes are divided into the sets $\{a_i\}$ and $\{b_i\}$, each of which is processed by an $N \times N$ quantum Fourier transform (QFT)~\cite{PhysRevLett.119.080502,PhysRevLett.73.58}
\begin{equation}
	\mathcal{F}(\hat{\alpha}_i^\dag)=\frac{1}{\sqrt{N}}\sum_{j=1}^N \ee^{\ii\ \frac{2  \pi}{N} ij}\hat{\alpha}^\dag_j,\quad\alpha=a,b
\end{equation}
thereby implementing global multimode mixing prior to photon-number-resolving detection. The probability of observing an output photon-number configuration $\vec{n}=(n_{a_1},n_{b_1},\dots,n_{a_N},n_{b_N})$ is completely determined by the unitary transformation describing the interferometric network~\cite{PhysRevLett.73.58,clements2016optimal}. For arbitrary input and output configurations, the transition amplitudes are given by the permanent of a submatrix drawn from the overall unitary, while the corresponding probabilities are obtained from their squared moduli. Evaluating the matrix permanent is computationally hard, nevertheless, in Appendix~\ref{nq}, we derive a general analytical expression for the output probabilities,
\begin{equation}
	P_{\vec{n}} \propto\left\vert\sum_{\vec{\sigma}\in S(N)}\exp\left(\ii \frac{2\pi}{N} \vec{\gamma}\cdot\vec{\sigma}+\ii\phi\mathcal{M}_{\mu,\sigma}\right)\right\vert^2.
    \label{eq:ProbMain}
\end{equation}
where the sum runs over the permutation group $S(N)$, and $\vec{\gamma}$ denotes an ordered sequence of output-port labels,
\begin{equation}
\vec{\gamma}
=
(\underbrace{1,\ldots,1}_{n_{1,a}},
\underbrace{1,\ldots,1}_{n_{1,b}},
\ldots,
\underbrace{N,\ldots,N}_{n_{N,a}},
\underbrace{N,\ldots,N}_{n_{N,b}})
\equiv
(\gamma_i)_{i=1,\ldots,N},
\end{equation}
whose entries identify the output port associated with each detected photon. The set
\begin{equation}
\mu=\{t\,|\,\text{the }t\text{th entry of }\vec{\gamma}\text{ corresponds to mode }b\},
\end{equation}
labels the photons detected in the $b$ modes, and the quantity $\mathcal M_{\mu,\sigma}$ counts how many of the indices $\sigma_t$, with $t\in\mu$, belong to the first $k$ input modes, which acquire the phase $\phi$. 
We can easily recognize the modulo-square permanent structure of Eq.~\eqref{eq:ProbMain} from the sum over permutations of the product of phases.
Further details and an illustrative example for the case $N=2$ are provided in Appendix~\ref{nq}.
Despite its complex form, the probability distribution $P_{\vec{n}}$ in Eq.~\eqref{eq:ProbMain} allows for an exact evaluation of the classical Fisher information (CFI) $F_{k,l}$ associated with the estimation of small phases $\phi\simeq0$. We find that non-vanishing contributions to the CFI originate only from a specific subset of the rare events $\vec{n}^*$, with zero probability, $P_{\vec{n}}^*=0$, that satisfy the condition
\begin{equation}
	\sum_{i=1}^N\gamma^*_i=0\ \text{mod }N.
    \label{eq:ConditionEvents}
\end{equation}
Interestingly, the condition in Eq.~\eqref{eq:ConditionEvents} depends on whether $N$ is even or odd. For example, events with one photon detected in each mode pair 
\(\{\hat{a}_i,\hat{b}_i\}_{i=1,\dots,N}\) are characterized by 
\(\gamma_i=i\), yielding  \(\sum_i \gamma_i = N(N-1)/2\), which is equal to $0$ mod $N$ only if $N$ is even. Therefore, these detection events provide useful information only when $N$ is even. Further details are provided in Appendix~\ref{nq}, where we show that, for $\phi\simeq 0$, the CFI is
\begin{equation}
F_{k,l} = \frac{kl}{N}\equiv N \bar{k}\bar{l},
\label{eq:Fisherkln}
\end{equation}
with $\bar{k}= k/N$ and $\bar{l}=l/N$ denoting the fractions of phase-encoding and reference photons, respectively. Writing $\bar{k}=(1+\Delta)/2$ and $\bar{l}=(1-\Delta)/2$ with the relative imbalance defined as $\Delta=|k-l|/N$, the CFI can be expressed as
\begin{equation}\nonumber
F_{k,l} = \frac{N}{4}\left(1 - \Delta^2\right).
\end{equation}
This shows that the CFI attains its maximum uniquely for a balanced partition, $\Delta=0$, and decreases quadratically with increasing imbalance. This highlights the fact that phase information arises from the correlation between the phase-encoding and reference sectors, the number of which scales as $kl$ and is maximized  for an equal partitioning of the photons. For arbitrary $\phi \neq 0$, the CFI generally exhibits a nontrivial dependence on $\phi$, as shown in \figurename~\ref{fig:FI2RB}. However, numerical results up to $N=6$ indicate that this dependence disappears for the symmetric partition $k=l$, rendering the CFI phase-independent.

\begin{table}
\centering
\begin{tabular}{p{1.30cm} p{1.30cm} p{1.30cm} p{1.30cm}} 
 \hline  
 \centering
 $N$ & $k$ & 
$l$ & $F_{k,l}$  \\
\hline  
\centering
  2 & 1 & 1 & $1/2$\\ 
 &  & &  \\
\hline
\centering
  3 & 1 & 2 & $2/3$\\ 
  \centering
   & 2 & 1 & $2/3$\\ 
 &  & &  \\
\hline
\centering
  &  & &  \\ 
  \centering
 4 & 1 & 3 & $3/4$\\ 
 \centering
   & 2 & 2 & $1$\\ 
   \centering
 & 3 & 1 & $3/4$\\ 
  &  & &  \\ 
    \hline
    \centering
  5 & 1 & 4 & $4/5$\\ 
  \centering
   & 2 & 3 & $6/5$\\
    \centering
   & 3 & 2 & $6/5$\\
    \centering
   & 4 & 1 & $4/5$\\
 &  & &  \\
\hline
   \centering
 6 & 1 & 5 &  $5/6$\\
 \centering
    & 2 & 4 &  $4/3$\\
     \centering
    & 3 & 3 &  $3/2$\\
     \centering
    & 4 & 2 &  $4/3$\\
     \centering
    & 5 & 1 &  $5/6$\\
  \hline  
\end{tabular}
\caption{The table reports the CFI for different input photon numbers $N$, evaluated numerically in the limit $\phi\rightarrow0$. The results show that increasing $N$ enhances the achievable phase sensitivity. For a fixed $N$, the CFI depends on the photon-number partition $(k,l)$, reaching its maximum for the balanced distribution $(k=l)$ and decreasing monotonically as the partition becomes increasingly asymmetric. An analytical proof of this sensitivity pattern is presented in the Appendix~\ref{nq}.}
\label{adm}
\end{table}

\vspace{0.5cm}  

\begin{table}\label{t}
\centering
\begin{tabular}{p{1.30cm} p{1.30cm} p{1.30cm} p{1.30cm}} 
 \hline  
 \centering
 $N$ & $k$ & 
$l$ & $F_{k,l}$  \\
\hline  
\centering
  2 & 1 & 1 & $\eta\epsilon/2$\\ 
&  & &  \\
\hline
\centering
  &  & &  \\ 
  \centering
 3 & 1 & 2 & $2\eta^2\epsilon/3$\\ 
 \centering
   & 2 & 1 & $2\eta\epsilon^2/3$\\ 
  &  & &  \\ 
    \hline
   \centering
 4 & 1 & 3 &  $3\eta^3\epsilon/4$\\
 \centering
    & 2 & 2 &  $\eta^2\epsilon^2$\\
     \centering
    & 3 & 1 &  $3\eta\epsilon^3/4$\\
   &  & &  \\   
  \hline  
\end{tabular}
\caption{
The table presents the CFI for photon-number sectors $(N,k,l)$ in the presence of loss, modeled by transmission efficiencies $\epsilon$ and $\eta$ for phase-encoded and reference photons, respectively. The FI for each distribution is evaluated in the limit $\phi \rightarrow 0$
With loss, and for $\phi =0$, each contribution is weighted by factors of $\eta$ and $\epsilon$, reflecting reduced transmission and detection probabilities. 
}
\label{adv}
\end{table}
\section{Lossy transmission}\label{inc}
\begin{figure}
    \centering
    \includegraphics[width=0.95\linewidth]{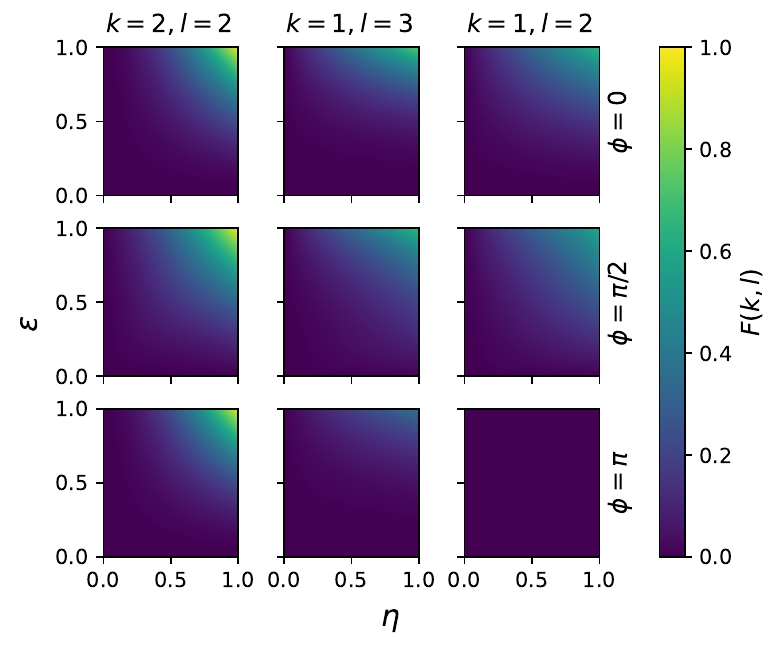}
    \includegraphics[width=0.95\linewidth]{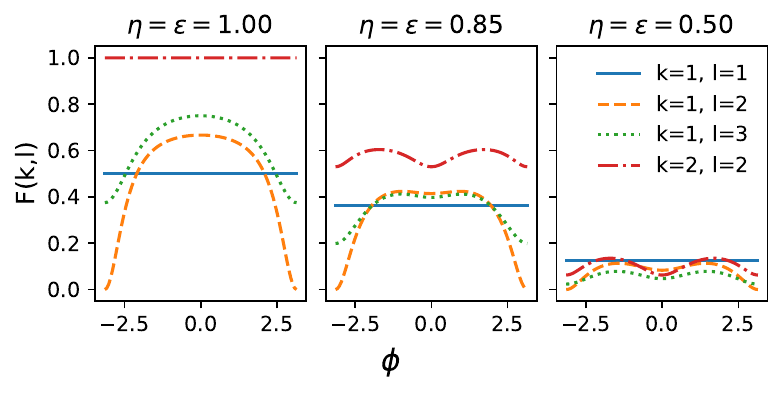}
    \caption{The table highlights the suppression of multiphoton advantage under loss. The CFI in a two-receiver configuration is shown in (a) as a function of the loss parameters $\epsilon$ and $\eta$ for representative values of $\phi$, $k$, and $l$, and (b) as a function of the phase $\phi$ for different loss regimes. Increasing loss progressively suppresses the multiphoton enhancement, leading to a systematic reduction in phase-estimation performance}
    \label{fig:FI2RB}
\end{figure}

Transmission loss is modeled by embedding the interferometric evolution into a quantum channel, where each optical mode is coupled to an independent environmental vacuum mode via a beam-splitter-like interaction~\cite{gardiner2004quantum}. We distinguish two transmissivities: $\epsilon$ for the phase-encoding modes and $\eta$ for the phase-independent reference modes. The corresponding Heisenberg-picture evolution of the annihilation operators is given by
\begin{align}
\hat{a}_j &\to \sqrt{\epsilon}\,\hat{a}_j + \sqrt{1-\epsilon}\,\hat{e}_j, \\
\hat{a}_k &\to \sqrt{\eta}\,\hat{a}_k + \sqrt{1-\eta}\,\hat{e}_k,
\end{align}
where $\hat{e}_{j,k}$ denotes the environmental vacuum operators. The overall evolution is thus described by the lossy channel $\mathcal{E}_{\epsilon,\eta} \circ \hat{U}$, which maps an initial pure $N$-photon state to a mixed state with  fluctuating photon number~\cite{10.1093/acprof:oso/9780199213900.001.0001}. Loss suppresses multiphoton interference and consequently degrades phase sensitivity~\cite{PhysRevLett.102.040403}. We report in Table~\ref{adv} the values of the CFI in the lossy scenario for $\phi \simeq 0$, computed by evaluating the output probabilities using the matrix permanent formalism. The same numerical approach is used to obtain the results for $\phi \neq 0$. Notably, the CFI values at $\phi \simeq 0$ can be obtained directly from the corresponding lossless values reported in Table~\ref{adm} by simply multiplying them by the factor $\eta^l \epsilon^k$. This follows from the fact that, in the $\phi \simeq 0$ limit, only events in which no photon is lost contribute to the CFI, yielding an overall scaling given by the probability that all photons are preserved, $\eta^l\epsilon^k$.
Away from $\phi \simeq 0$, however, the dependence on the loss parameters $\epsilon$ and $\eta$ becomes considerably more intricate. In this regime, the CFI receives contributions from all possible loss channels, leading to a nontrivial combinatorial structure that does not admit a simple closed-form expression. This behavior is illustrated in Fig.~\ref{fig:FI2RB}, where the CFI is shown as a function of the loss parameters $\epsilon$ and $\eta$ for representative values of $\phi$, $k$, and $l$, together with its dependence on the phase $\phi$ for different loss regimes. In all cases, increasing losses progressively suppress the multiphoton enhancement, leading to a monotonic degradation of the phase-estimation performance.

\section{Comparison with local homodyne detection}\label{ind}
The phase estimation protocol based on single-photon probe typically typically requires a reference photon to retrieve the encoded phase. In practice, however,  transmission losses can degrade this reference and limit precision. This motivates exploring whether a classical local oscillator combined with homodyne detection offers a more robust alternative. We therefore consider the single-photon state
\begin{align}
\ket{\psi}=\frac{1}{\sqrt{2}}\left(\hat{a}^\dagger + e^{\ii\phi} \hat{b}^\dagger\right)\ket{0},
\end{align}
and assume access to a shared coherent phase reference, allowing local quadrature measurements on each mode. The quadrature operator for mode $\alpha\in\{a,b\}$ is defined as
\begin{align}
\hat{x}_\alpha(\theta)=\frac{1}{\sqrt{2}}\left(\hat{\alpha}\,e^{-i\theta}+\hat{\alpha}^\dagger e^{\ii\theta}\right),
\end{align}
where $\theta$ is the local oscillator phase. We evaluate the phase sensitivity of homodyne detection using the CFI associated with measurements of \(\hat{x}_a(0)\equiv\hat{x}_a\) and \(\hat{x}_b(\pi/2)\equiv\hat{p}_b\). Introducing the rotated mode
\begin{align}
\hat{c}^\dagger
=
\frac{1}{\sqrt{2}}
\left(
\hat{a}^\dagger
+
e^{\ii\phi} \hat{b}^\dagger
\right),
\end{align}
the single-photon state can be written as $\ket{\psi}=\hat{c}^\dagger\ket{0}$.
In Appendix~\ref{mx} we employ the Wigner distribution of the single-photon state to reconstruct the joint probability distribution 
\begin{equation}
P(x_a,p_b)=\frac{1}{\pi}e^{-x_a^2-p_b^2}
\left(x_a^2+p_b^2+2x_ap_b\sin\phi\right).
\label{cz}
\end{equation}
The corresponding continuous-variable CFI,
\begin{align}
F_H(\phi)=\int dx_a\,dp_b\,
\frac{\left[\partial_\phi P(x_a,p_b)\right]^2}{P(x_a,p_b)},
\end{align}
captures the total phase information accessible through homodyne detection. We find that $F_H(\phi)$ is maximized at $\phi=0$, yielding
\begin{align}
F_H=\frac{1}{2}.
\end{align}
However, the sensitivity detoriates in presence of loss where the vacuum noise mixes with the probability in Eq.~\eqref{cz}, obtaining (see Appendix~\ref{mx})
\begin{equation}
    P_\varepsilon(x_a,p_b)=\frac{1}{\pi}e^{-x_a^2-p_b^2}\left[1-\varepsilon+\varepsilon\left(x_a^2+p_b^2+2x_a p_b\sin\phi\right)\right].
\end{equation}
In the low-transmission regime, \(\epsilon\ll1\), the corresponding continuous-variable CFI scales as
\begin{align}
F_{H,\epsilon}\sim \epsilon^2\cos^2\phi,
\end{align}
consistent with the bounds of Ref.~\cite{PhysRevLett.107.270402}. In contrast, photon-number-resolving detection in the same interferometric scheme yields a CFI that scales linearly with the transmission,
\begin{align}\label{my}
F_{1,1}=\frac{\epsilon}{2},
\end{align}
as summarized in Table~\ref{t}, where $F_{1,1}$ denotes the CFI obtained using a single phase-encoding photon and a single reference photon. This improvement arises because detection events carrying no phase information can be discarded without reducing the information obtained from successful trials. Therefore, photon-number-resolving detection avoids the quadratic loss scaling of homodyne detection and remains substantially more efficient in the low-transmission regime. This linear loss scaling is specific to the single phase-encoding scenario considered here. More generally, for a probe containing \(k\) phase-encoding photons, successful estimation requires the transmission of all encoded photons, leading to a CFI that typically scales as \(\epsilon^{k}\).

\begin{figure*}\label{rec}
\centering\includegraphics[width=.73\linewidth,height=2.9in]{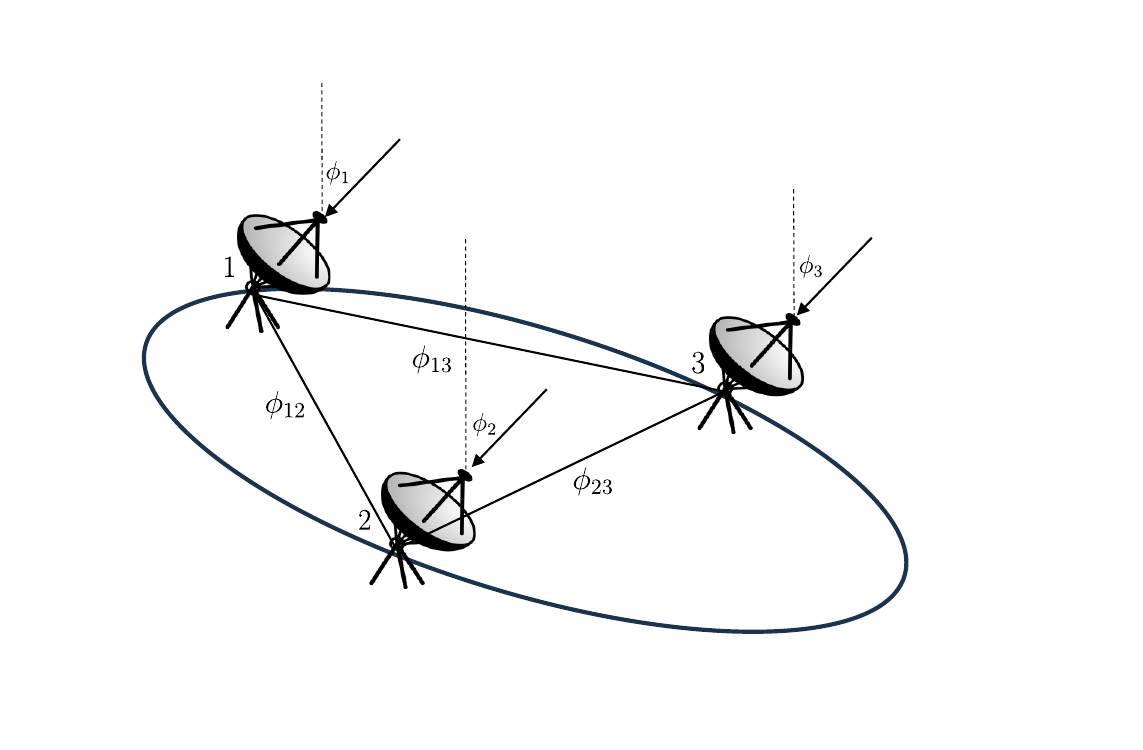}
\caption{Schematic of a three-receiver single-photon optical telescope for multiparameter phase estimation. A single photon is coherently distributed among three spatial modes associated with the three receivers. The relative phases between the modes are encoded through the optical path differences between the source and the corresponding receivers, with one mode chosen as an arbitrary phase reference to remove the global phase freedom. The resulting probe state contains the relevant relative phase information and is characterised by a full-rank quantum Fisher information matrix. The three pairwise baselines provide multiparameter estimation and precision analysis via the quantum Cramér–Rao bound.
}
\end{figure*}

\section{From Two-Mode to MultiMode Single-Photon Interferometry}\label{ine}

\begin{figure}
\centering
\includegraphics[width=0.95\linewidth]{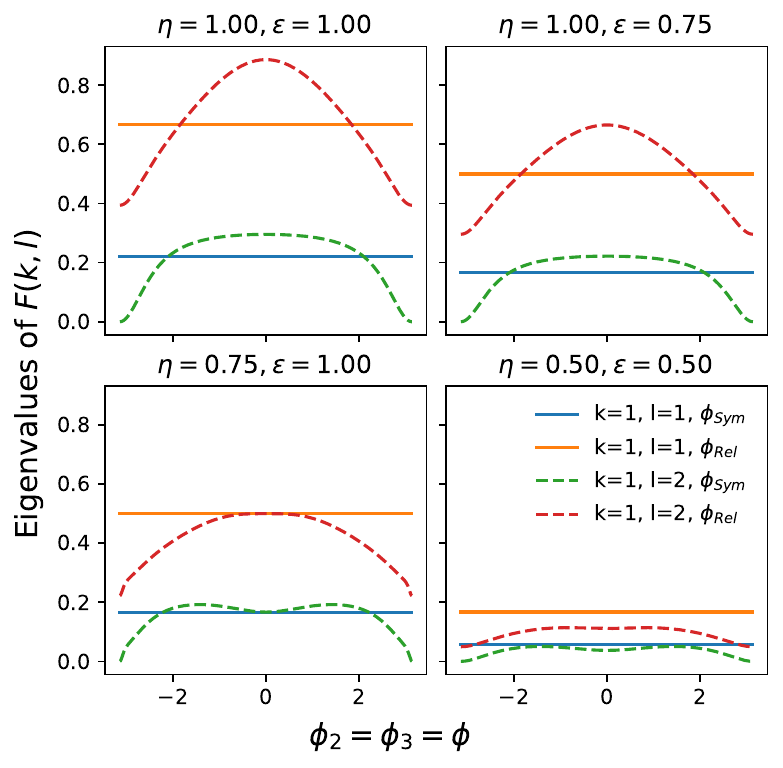}
\caption{Illustrates CFI under transmission loss. The eigenvalues of the CFI matrix are evaluated for $N$-photon state undergoing lossy evolution, with $\epsilon$ and $\eta$ denoting the transmissivities of the phase and reference modes, respectively. Results are shown for $N=2$ and $N=3$, corresponding to the photon-number partitions $(k,l)=(1,1)$ and $(1,2)$ under the constraint $\phi_2=\phi_3$. Increasing the number of reference photons initially improves the precision of relative-phase estimation. However, this improvement persists only up to a critical balance between $\eta$ and $\epsilon$. Beyond this point, the metrological advantage is lost, resulting in reduced sensitivity.} 
\label{cq}
\end{figure}
We consider a single photon distributed over $M$ modes and fix the global phase by choosing the first mode as the reference. The probe state is
\begin{align}\label{sz}
\ket{\psi}=\frac{1}{\sqrt{M}}\left(\ket{1}+\sum_{j=2}^{M} e^{\ii\phi_j}\ket{j}\right),
\end{align}
where $\boldsymbol{\phi}=(\phi_2,\ldots,\phi_M)$
denotes the $(M-1)$ independent phases. The generator associated with $\varphi_j$ is the local number operator $G_j = \hat{a}_j^\dagger \hat{a}_j$, which commutes and satisfies
\begin{align}
G_j\ket{\psi}=\frac{1}{\sqrt{M}} e^{\ii\phi_j}\ket{j}.
\end{align}
For pure states and commuting generators, the quantum Fisher information matrix (QFIM) is given by the covariance matrix of the generators \cite{doi:10.1142/S0219749909004839},
\begin{align}
(Q_{\phi})_{jk}
=4\left(\langle G_j G_k\rangle - \langle G_j\rangle \langle G_k\rangle\right)
=4\left(\frac{\delta_{jk}}{M}-\frac{1}{M^2}\right).
\end{align}
This can be written compactly as
\begin{align}
Q_{\phi}=\frac{4}{M}\left(I_{M-1}-\frac{1}{M}\mathbf{1}\mathbf{1}^\top\right),
\end{align}
where $\mathbf{1}=(1,\dots,1)^\top$ denotes the $(M-1)$-dimensional vector whose entries are all unity, and $\mathbf{1}\mathbf{1}^\top$ is the rank-1 projector onto the symmetric direction.
The spectrum of $Q_{\phi}$ has a nondegenerate eigenvalue $\lambda_0=4/M^2$, associated with the symmetric vector $\mathbf{1}$, and degenerate eigenvalue $\lambda=4/M$ corresponding to the subspace orthogonal to $\mathbf{1}$. The QFIM eigenvectors identify the optimal parameter combinations, while the corresponding eigenvalues determine their attainable precision. For $M=3$, it is convenient to introduce the orthonormal combinations
\begin{equation}
\phi_{\rm Rel}
=
\frac{\phi_2-\phi_3}{\sqrt{2}},
\qquad
\phi_{\rm Sym}
=
\frac{\phi_2+\phi_3}{\sqrt{2}},
\end{equation}
which coincide with the eigenvectors of the QFIM. The corresponding quantum Cramér-Rao bounds are
\begin{align}
\mathrm{Var}(\phi_{\mathrm{Rel}}) \ge \frac{3}{4}, 
\qquad
\mathrm{Var}(\phi_{\mathrm{Sym}}) \ge \frac{9}{4}.
\end{align}
Thus, the symmetric phase combination is estimated with lower precision than the relative phase combinations. More generally, defining
\(\operatorname{Var}_{\mathrm{Sym}}=\lambda_0^{-1}\) and
\(\operatorname{Var}_{\mathrm{Rel}}=\lambda^{-1}\), we obtain
\begin{equation}
\frac{\operatorname{Var}_{\mathrm{Sym}}}{\operatorname{Var}_{\mathrm{Rel}}}
= M.
\end{equation}
This indicates that the estimation uncertainty of the symmetric mode is always greater than that of the relative modes for all values of $M$.
An intuitive explanation for the enhanced sensitivity of the relative-phase mode is provided in Appendix~\ref{zl}. Motivated by this structure, we interpret the three modes as receivers in an optical interferometric array, where additional receivers increase the number of available baselines and modify the accessible phase correlations. To quantify the advantage provided by additional baselines, we compare the CFI sensitivities for different photon partitions, as summarized in Table~\ref{adv3}. Interestingly, numerical verification for \(N \leq 4\) shows that, for each mode pair \((k,l)\), the CFI at the zero-phase point \(\phi \approx 0\) retains the same functional form as in the two-mode case, differing only by an overall scaling factor
\begin{equation}
F_{k,l}(\phi_{\mathrm{Rel}})=\frac{4}{3}F_{k,l}, \qquad
F_{k,l}(\phi_{\mathrm{Sym}})=\frac{4}{9}F_{k,l},
\end{equation}
where \(F_{k,l}=kl/N\). This correspondence remains valid under transmission loss at the zero-phase point, with loss modifying only the prefactors. Furthermore, for a fixed photon number \(N\), the maximum sensitivity is achieved with a balanced photon distribution. The extension of these results to arbitrary $M$ and $N$ is yet to be investigated. We further find that the relative mode consistently yields a lower estimation variance than the symmetric mode, revealing the anisotropic distribution of information and the gap between experimentally accessible measurements and the quantum limit.

The effect of transmission loss is incorporated using the model introduced in 
Sec.~\ref{inc}, with the corresponding results presented in Fig.~\ref{cq}. We numerically evaluate the CFI for arbitrary photon numbers $N$ under lossy evolution. For $\phi_2=\phi_3$, increasing the number of reference photons initially improves relative-phase sensitivity. This advantage, however, persists only below a critical imbalance between the transmissivities $\eta$ and $\epsilon$, beyond which the estimation deteriorates. Next, we examine phase estimation based on local homodyne detection and characterize its sensitivity to optical loss. In the high photon-loss regime ($\epsilon\ll1$), vacuum contributions dominate the measurement statistics, leading to the quadratic CFI scaling
\begin{equation}
F_{\epsilon}(\phi)\propto \epsilon^{2},
\end{equation}
as derived in Appendix~\ref{pz}. This contrasts with the linear scaling achievable with global measurements discussed in Sec.~\ref{ind}. Local homodyne detection is therefore suboptimal for highly attenuated delocalized single-photon states, as optical loss degrades the intermode coherence responsible for phase sensitivity.
\vspace{1em}
\begin{table}
\centering
\begin{tabular}{p{1.30cm} p{1.30cm} p{1.30cm} p{1.30cm}  p{1.30cm}} 
 \hline  
 \centering
 $N$ & $k$ & 
$l$ & $\mathrm{Var}(\phi_{\mathrm{Rel}})$ & $\mathrm{Var}(\phi_{\mathrm{Sym}})$  \\
\hline  
\centering
  2 & 1 & 1 & $3\eta\epsilon/2$ & $9\eta\epsilon/2$\\ 
&  & &  \\
\hline
\centering
  &  & &  \\ 
  \centering
 3 & 1 & 2 & $9\eta^2\epsilon/8$ & $27\eta^2\epsilon/8$\\ 
 \centering
   & 2 & 1 & $9\eta\epsilon^2/8$ & $27\eta\epsilon^2/8$\\ 
  &  & &  \\ 
    \hline
\centering
  &  & &  \\ 
  \centering
 4 & 1 & 3 & $\eta^3\epsilon$ & $3\eta^3\epsilon$\\ 
 \centering
   & 2 & 2 & $3\eta^2\epsilon^2/4$ & $9\eta^2\epsilon^2/4$\\ 
  \centering
   & 3 & 1 & $\eta\epsilon^3$ & $3\eta\epsilon^3$\\ 
  &  & &  \\ 
    \hline
\end{tabular}
\caption{The table presents the CFI-based lower bounds on phase-estimation variances for different photon partitions $(k,l)$ in a three-mode interferometer under transmission loss. The results are evaluated numerically at the zero-phase point $\phi \simeq 0$. As in the two-mode scenario, the highest sensitivity is achieved for a balanced photon distribution. Across all configurations, the relative mode consistently outperforms the symmetric mode, with precision improving as the total photon number increases.}
\label{adv3}
\end{table}
\section{Conclusions}\label{ss}
We have investigated linear-optical interferometric protocols in which separable multiphoton inputs acquire number-path entanglement through passive linear optics. By deriving the output statistics, we obtain a closed-form expression for the CFI at the zero-phase point. A balanced partition maximizes the sensitivity for arbitrary photon numbers, whereas asymmetric configurations lead to reduced performance. We further analyze the impact of loss and find that, at zero phase, the CFI is suppressed only by the probability of transmitting all photons, while away from this point, multiple loss channels give rise to more complex behavior. Comparing measurement schemes, we show that photon-number-resolving detection retains linear multiphoton scaling, whereas local homodyne detection is limited by vacuum fluctuations. Extending the framework to distributed single-photon interferometry with multiple receivers, we find that relative-phase modes carry the dominant metrological information, while symmetric modes become increasingly difficult to estimate in higher-dimensional multimode networks. These results reveal a direct connection between the geometry of distributed interferometric architectures and their attainable precision.

Our findings provide design principles for scalable quantum sensing platforms based on passive linear optics, identifying optimal photon allocations and measurement strategies for distributed interferometry and quantum-enhanced imaging.

\section*{Acknowledgements}\noindent

This work has received funding from the
European Union's Horizon Europe research and innovation programme under the Project ``Quantum Secure Networks Partnership'' (QSNP, Grant Agreement No.~101114043);
and from the Italian Space Agency (Subdiffraction
Quantum Imaging ``SQI'' 2023-13-HH.0).

\bibliography{error}

@PREAMBLE{
 "\providecommand{\noopsort}[1]{}" 
 # "\providecommand{\singleletter}[1]{#1}%" 
}

@article{giovannetti2004quantum,
  title={Quantum-enhanced measurements: beating the standard quantum limit},
  author={Giovannetti, Vittorio and Lloyd, Seth and Maccone, Lorenzo},
  journal={Science},
  volume={306},
  number={5700},
  pages={1330--1336},
  year={2004},
  publisher={American Association for the Advancement of Science}
}

@article{dowling2008quantum,
  title={Quantum optical metrology--the lowdown on high-N00N states},
  author={Dowling, Jonathan P},
  journal={Contemporary physics},
  volume={49},
  number={2},
  pages={125--143},
  year={2008},
  publisher={Taylor \& Francis}
}

@article{helstrom1969quantum,
  title={Quantum detection and estimation theory},
  author={Helstrom, Carl W},
  journal={Journal of statistical physics},
  volume={1},
  number={2},
  pages={231--252},
  year={1969},
  publisher={Springer}
}

@article{walther2004broglie,
  title={De Broglie wavelength of a non-local four-photon state},
  author={Walther, Philip and Pan, Jian-Wei and Aspelmeyer, Markus and Ursin, Rupert and Gasparoni, Sara and Zeilinger, Anton},
  journal={Nature},
  volume={429},
  number={6988},
  pages={158--161},
  year={2004},
  publisher={Nature Publishing Group UK London}
}

@article{clements2016optimal,
  title={Optimal design for universal multiport interferometers},
  author={Clements, William R and Humphreys, Peter C and Metcalf, Benjamin J and Kolthammer, W Steven and Walmsley, Ian A},
  journal={Optica},
  volume={3},
  number={12},
  pages={1460--1465},
  year={2016},
  publisher={Optical Society of America}
}

@article{PhysRevLett.102.100401,
  title = {Entanglement, Nonlinear Dynamics, and the Heisenberg Limit},
  author = {Pezz\'e, Luca and Smerzi, Augusto},
  journal = {Phys. Rev. Lett.},
  volume = {102},
  issue = {10},
  pages = {100401},
  numpages = {4},
  year = {2009},
  month = {Mar},
  publisher = {American Physical Society},
  doi = {10.1103/PhysRevLett.102.100401},
  url = {https://link.aps.org/doi/10.1103/PhysRevLett.102.100401}
}

@article{doi:10.1142/S0219749909004839,
author = {Paris, MATTEO G. A.},
title = {QUANTUM ESTIMATION FOR QUANTUM TECHNOLOGY},
journal = {International Journal of Quantum Information},
volume = {07},
number = {supp01},
pages = {125-137},
year = {2009},
doi = {10.1142/S0219749909004839}
}

@article{KIM199837,
title = {The phase-sensitivity of a Mach–Zehnder interferometer for Fock state inputs},
journal = {Optics Communications},
volume = {156},
number = {1},
pages = {37-42},
year = {1998},
issn = {0030-4018},
doi = {https://doi.org/10.1016/S0030-4018(98)00428-3},
url = {https://www.sciencedirect.com/science/article/pii/S0030401898004283},
author = {Taesoo Kim and Jongtae Shin and Yang Ha and Heonoh Kim and Goodong Park and Tae Gon Noh and Chung Ki Hong}
}

@article{PhysRevD.23.1693,
  title = {Quantum-mechanical noise in an interferometer},
  author = {Caves, Carlton M.},
  journal = {Phys. Rev. D},
  volume = {23},
  issue = {8},
  pages = {1693--1708},
  numpages = {0},
  year = {1981},
  month = {Apr},
  publisher = {American Physical Society},
  doi = {10.1103/PhysRevD.23.1693},
  url = {https://link.aps.org/doi/10.1103/PhysRevD.23.1693}
}

@article{PhysRevA.33.4033,
  title = {SU(2) and SU(1,1) interferometers},
  author = {Yurke, Bernard and McCall, Samuel L. and Klauder, John R.},
  journal = {Phys. Rev. A},
  volume = {33},
  issue = {6},
  pages = {4033--4054},
  numpages = {0},
  year = {1986},
  month = {Jun},
  publisher = {American Physical Society},
  doi = {10.1103/PhysRevA.33.4033},
  url = {https://link.aps.org/doi/10.1103/PhysRevA.33.4033}
}

@article{
doi:10.1126/science.1138007,
author = {Tomohisa Nagata  and Ryo Okamoto  and Jeremy L. O'Brien  and Keiji Sasaki  and Shigeki Takeuchi },
title = {Beating the Standard Quantum Limit with Four-Entangled Photons},
journal = {Science},
volume = {316},
number = {5825},
pages = {726-729},
year = {2007},
doi = {10.1126/science.1138007}}

@article{PhysRevA.80.043822,
  title = {Resolution and sensitivity of a Fabry-Perot interferometer with a photon-number-resolving detector},
  author = {Wildfeuer, Christoph F. and Pearlman, Aaron J. and Chen, Jun and Fan, Jingyun and Migdall, Alan and Dowling, Jonathan P.},
  journal = {Phys. Rev. A},
  volume = {80},
  issue = {4},
  pages = {043822},
  numpages = {7},
  year = {2009},
  month = {Oct},
  publisher = {American Physical Society},
  doi = {10.1103/PhysRevA.80.043822},
  url = {https://link.aps.org/doi/10.1103/PhysRevA.80.043822}
}

@article{PhysRevLett.119.080502,
  title = {Multiphoton Interference in Quantum Fourier Transform Circuits and Applications to Quantum Metrology},
  author = {Su, Zu-En and Li, Yuan and Rohde, Peter P. and Huang, He-Liang and Wang, Xi-Lin and Li, Li and Liu, Nai-Le and Dowling, Jonathan P. and Lu, Chao-Yang and Pan, Jian-Wei},
  journal = {Phys. Rev. Lett.},
  volume = {119},
  issue = {8},
  pages = {080502},
  numpages = {5},
  year = {2017},
  month = {Aug},
  publisher = {American Physical Society},
  doi = {10.1103/PhysRevLett.119.080502},
  url = {https://link.aps.org/doi/10.1103/PhysRevLett.119.080502}
}

@article{PhysRevLett.71.1355,
  title = {Interferometric detection of optical phase shifts at the Heisenberg limit},
  author = {Holland, M. J. and Burnett, K.},
  journal = {Phys. Rev. Lett.},
  volume = {71},
  issue = {9},
  pages = {1355--1358},
  numpages = {0},
  year = {1993},
  month = {Aug},
  publisher = {American Physical Society},
  doi = {10.1103/PhysRevLett.71.1355},
  url = {https://link.aps.org/doi/10.1103/PhysRevLett.71.1355}
}

@article{RevModPhys.90.035005,
  title = {Quantum metrology with nonclassical states of atomic ensembles},
  author = {Pezz\`e, Luca and Smerzi, Augusto and Oberthaler, Markus K. and Schmied, Roman and Treutlein, Philipp},
  journal = {Rev. Mod. Phys.},
  volume = {90},
  issue = {3},
  pages = {035005},
  numpages = {70},
  year = {2018},
  month = {Sep},
  publisher = {American Physical Society},
  doi = {10.1103/RevModPhys.90.035005},
  url = {https://link.aps.org/doi/10.1103/RevModPhys.90.035005}
}

@article{PhysRevLett.85.2733,
  title = {Quantum Interferometric Optical Lithography: Exploiting Entanglement to Beat the Diffraction Limit},
  author = {Boto, Agedi N. and Kok, Pieter and Abrams, Daniel S. and Braunstein, Samuel L. and Williams, Colin P. and Dowling, Jonathan P.},
  journal = {Phys. Rev. Lett.},
  volume = {85},
  issue = {13},
  pages = {2733--2736},
  numpages = {0},
  year = {2000},
  month = {Sep},
  publisher = {American Physical Society},
  doi = {10.1103/PhysRevLett.85.2733},
  url = {https://link.aps.org/doi/10.1103/PhysRevLett.85.2733}
}

@article{PhysRevLett.107.270402,
  title = {Quantum Nonlocality in Weak-Thermal-Light Interferometry},
  author = {Tsang, Mankei},
  journal = {Phys. Rev. Lett.},
  volume = {107},
  issue = {27},
  pages = {270402},
  numpages = {5},
  year = {2011},
  month = {Dec},
  publisher = {American Physical Society},
  doi = {10.1103/PhysRevLett.107.270402},
  url = {https://link.aps.org/doi/10.1103/PhysRevLett.107.270402}
}

@article{Huang31122026,
author = {Zixin Huang and Oleg Titov and Mikołaj K. Schmidt and Benjamin Pope and Gavin K. Brennen and Daniel K. L. Oi and Pieter Kok},
title = {Quantum-enabled optical large-baseline interferometry: applications, protocols and feasibility},
journal = {Advances in Physics: X},
volume = {11},
number = {1},
pages = {2597311},
year = {2026},
publisher = {Taylor \& Francis},
doi = {10.1080/23746149.2025.2597311},
}

@article{10.1098/rspa.2011.0232,
    author = {Aaronson, Scott},
    title = {A linear-optical proof that the permanent is P-hard},
    journal = {Proceedings of the Royal Society A: Mathematical, Physical and Engineering Sciences},
    volume = {467},
    number = {2136},
    pages = {3393-3405},
    year = {2011},
    month = {07},
    issn = {1364-5021},
    doi = {10.1098/rspa.2011.0232}
}

@article{PhysRevA.108.052408,
  title = {Optimal qubit circuits for quantum-enhanced telescopes},
  author = {Czupryniak, Robert and Steinmetz, John and Kwiat, Paul G. and Jordan, Andrew N.},
  journal = {Phys. Rev. A},
  volume = {108},
  issue = {5},
  pages = {052408},
  numpages = {13},
  year = {2023},
  month = {Nov},
  publisher = {American Physical Society},
  doi = {10.1103/PhysRevA.108.052408},
  url = {https://link.aps.org/doi/10.1103/PhysRevA.108.052408}
}

@article{354q-ch63,
  title = {Superresolution Imaging with Entanglement-Enhanced Telescopy},
  author = {Padilla, Isack and Sajjad, Aqil and Saif, Babak N. and Guha, Saikat},
  journal = {Phys. Rev. Lett.},
  volume = {136},
  issue = {1},
  pages = {010803},
  numpages = {7},
  year = {2026},
  month = {Jan},
  publisher = {American Physical Society},
  doi = {10.1103/354q-ch63},
  url = {https://link.aps.org/doi/10.1103/354q-ch63}
}

@article{PhysRevLett.131.210801,
  title = {Interferometric Imaging Using Shared Quantum Entanglement},
  author = {Brown, Matthew R. and Allgaier, Markus and Thiel, Val\'erian and Monnier, John D. and Raymer, Michael G. and Smith, Brian J.},
  journal = {Phys. Rev. Lett.},
  volume = {131},
  issue = {21},
  pages = {210801},
  numpages = {6},
  year = {2023},
  month = {Nov},
  publisher = {American Physical Society},
  doi = {10.1103/PhysRevLett.131.210801},
  url = {https://link.aps.org/doi/10.1103/PhysRevLett.131.210801}
}

@article{hong2021quantum,
  title={Quantum enhanced multiple-phase estimation with multi-mode n 00 n states},
  author={Hong, Seongjin and ur Rehman, Junaid and Kim, Yong-Su and Cho, Young-Wook and Lee, Seung-Woo and Jung, Hojoong and Moon, Sung and Han, Sang-Wook and Lim, Hyang-Tag},
  journal={Nature communications},
  volume={12},
  number={1},
  pages={5211},
  year={2021},
  publisher={Nature Publishing Group UK London}
}

@article{PhysRevLett.96.010401,
  title = {Quantum Metrology},
  author = {Giovannetti, Vittorio and Lloyd, Seth and Maccone, Lorenzo},
  journal = {Phys. Rev. Lett.},
  volume = {96},
  issue = {1},
  pages = {010401},
  numpages = {4},
  year = {2006},
  month = {Jan},
  publisher = {American Physical Society},
  doi = {10.1103/PhysRevLett.96.010401},
  url = {https://link.aps.org/doi/10.1103/PhysRevLett.96.010401}
}

@article{Dowling01032008,
author = {Jonathan P. Dowling},
title = {Quantum optical metrology – the lowdown on high-N00N states},
journal = {Contemporary Physics},
volume = {49},
number = {2},
pages = {125--143},
year = {2008},
publisher = {Taylor \& Francis},
doi = {10.1080/00107510802091298},
}

@article{RevModPhys.79.135,
  title = {Linear optical quantum computing with photonic qubits},
  author = {Kok, Pieter and Munro, W. J. and Nemoto, Kae and Ralph, T. C. and Dowling, Jonathan P. and Milburn, G. J.},
  journal = {Rev. Mod. Phys.},
  volume = {79},
  issue = {1},
  pages = {135--174},
  numpages = {0},
  year = {2007},
  month = {Jan},
  publisher = {American Physical Society},
  doi = {10.1103/RevModPhys.79.135},
  url = {https://link.aps.org/doi/10.1103/RevModPhys.79.135}
}

@article{PhysRevLett.106.153603,
  title = {Optical Phase Estimation in the Presence of Phase Diffusion},
  author = {Genoni, Marco G. and Olivares, Stefano and Paris, Matteo G. A.},
  journal = {Phys. Rev. Lett.},
  volume = {106},
  issue = {15},
  pages = {153603},
  numpages = {4},
  year = {2011},
  month = {Apr},
  publisher = {American Physical Society},
  doi = {10.1103/PhysRevLett.106.153603},
  url = {https://link.aps.org/doi/10.1103/PhysRevLett.106.153603}
}

@article{PhysRevLett.73.58,
  title = {Experimental realization of any discrete unitary operator},
  author = {Reck, Michael and Zeilinger, Anton and Bernstein, Herbert J. and Bertani, Philip},
  journal = {Phys. Rev. Lett.},
  volume = {73},
  issue = {1},
  pages = {58--61},
  numpages = {0},
  year = {1994},
  month = {Jul},
  publisher = {American Physical Society},
  doi = {10.1103/PhysRevLett.73.58},
  url = {https://link.aps.org/doi/10.1103/PhysRevLett.73.58}
}

@article{PhysRevLett.100.073601,
  title = {Mach-Zehnder Interferometry at the Heisenberg Limit with Coherent and Squeezed-Vacuum Light},
  author = {Pezz\'e, Luca and Smerzi, Augusto},
  journal = {Phys. Rev. Lett.},
  volume = {100},
  issue = {7},
  pages = {073601},
  numpages = {4},
  year = {2008},
  month = {Feb},
  publisher = {American Physical Society},
  doi = {10.1103/PhysRevLett.100.073601},
  url = {https://link.aps.org/doi/10.1103/PhysRevLett.100.073601}
}

@article{PhysRevA.109.053508,
  title = {Tunable linear-optical phase amplification},
  author = {Schwarze, Christopher R. and Simon, David S. and Ndao, Abdoulaye and Sergienko, Alexander V.},
  journal = {Phys. Rev. A},
  volume = {109},
  issue = {5},
  pages = {053508},
  numpages = {6},
  year = {2024},
  month = {May},
  publisher = {American Physical Society},
  doi = {10.1103/PhysRevA.109.053508},
  url = {https://link.aps.org/doi/10.1103/PhysRevA.109.053508}
}

@book{10.1093/acprof:oso/9780199213900.001.0001,
    author = {Breuer, Heinz-Peter and Petruccione, Francesco},
    title = {The Theory of Open Quantum Systems},
    publisher = {Oxford University Press},
    year = {2007},
    month = {01},
    isbn = {9780199213900},
    doi = {10.1093/acprof:oso/9780199213900.001.0001},
    url = {https://doi.org/10.1093/acprof:oso/9780199213900.001.0001},
}

@book{gardiner2004quantum,
  title={Quantum noise: a handbook of Markovian and non-Markovian quantum stochastic methods with applications to quantum optics},
  author={Gardiner, Crispin and Zoller, Peter},
  year={2004},
  publisher={Springer Science \& Business Media}
}

@article{PhysRevLett.102.040403,
  title = {Optimal Quantum Phase Estimation},
  author = {Dorner, U. and Demkowicz-Dobrzanski, R. and Smith, B. J. and Lundeen, J. S. and Wasilewski, W. and Banaszek, K. and Walmsley, I. A.},
  journal = {Phys. Rev. Lett.},
  volume = {102},
  issue = {4},
  pages = {040403},
  numpages = {4},
  year = {2009},
  month = {Jan},
  publisher = {American Physical Society},
  doi = {10.1103/PhysRevLett.102.040403},
  url = {https://link.aps.org/doi/10.1103/PhysRevLett.102.040403}
}

@book{nielsen2010quantum,
  title={Quantum computation and quantum information},
  author={Nielsen, Michael A and Chuang, Isaac L},
  year={2010},
  publisher={Cambridge university press}
}

@article{tillmann2013experimental,
  title={Experimental boson sampling},
  author={Tillmann, Max and Daki{\'c}, Borivoje and Heilmann, Ren{\'e} and Nolte, Stefan and Szameit, Alexander and Walther, Philip},
  journal={Nature photonics},
  volume={7},
  number={7},
  pages={540--544},
  year={2013},
  publisher={Nature Publishing Group UK London}
}

@article{escher2011general,
  title={General framework for estimating the ultimate precision limit in noisy quantum-enhanced metrology},
  author={Escher, BM and de Matos Filho, Ruynet Lima and Davidovich, Luiz},
  journal={Nature Physics},
  volume={7},
  number={5},
  pages={406--411},
  year={2011},
  publisher={Nature Publishing Group UK London}
}

@article{demkowicz2015quantum,
  title={Quantum limits in optical interferometry},
  author={Demkowicz-Dobrza{\'n}ski, Rafal and Jarzyna, Marcin and Ko{\l}ody{\'n}ski, Jan},
  journal={Progress in Optics},
  volume={60},
  pages={345--435},
  year={2015},
  publisher={Elsevier}
}

@article{demkowicz2012elusive,
  title={The elusive Heisenberg limit in quantum-enhanced metrology},
  author={Demkowicz-Dobrza{\'n}ski, Rafa{\l} and Ko{\l}ody{\'n}ski, Jan and Gu{\c{t}}{\u{a}}, M{\u{a}}d{\u{a}}lin},
  journal={Nature communications},
  volume={3},
  number={1},
  pages={1063},
  year={2012},
  publisher={Nature Publishing Group UK London}
}

@article{nagata2007beating,
  title={Beating the standard quantum limit with four-entangled photons},
  author={Nagata, Tomohisa and Okamoto, Ryo and O'brien, Jeremy L and Sasaki, Keiji and Takeuchi, Shigeki},
  journal={Science},
  volume={316},
  number={5825},
  pages={726--729},
  year={2007},
  publisher={American Association for the Advancement of Science}
}

@inproceedings{aaronson2011computational,
  title={The computational complexity of linear optics},
  author={Aaronson, Scott and Arkhipov, Alex},
  booktitle={Proceedings of the forty-third annual ACM symposium on Theory of computing},
  pages={333--342},
  year={2011}
}
\widetext
\appendix\label{zz}

\section{Derivation of the Generalized Classical Fisher Information}\label{nq}
Here, we derive a general expression for the CFI at $\phi \simeq 0$ associated with the detection statistics of the interferometric scheme. We first express the probability amplitude in terms of the output-mode pattern and the subset of photons that undergo the phase shift. Next, we show that the linear term in the small-phase expansion vanishes for every detection outcome, implying that only events with zero probability at $\phi = 0$ can contribute to the CFI. Finally, we identify the outcomes that yield a nonzero contribution, leading to a closed-form expression for the CFI in the limit $\phi \simeq 0$.
\subsection{Characterization of Detection Outcomes and Their Phase Dependence}
We consider an interferometric scheme where each photon is distributed over a pair of optical modes $\{a_i,b_i\}$, with the first $k$ mode pairs acquiring the unknown phase $\phi$. We define $\phi_i$ as the phase applied to the $i$th mode pair
\begin{equation}
\phi_i =
\begin{cases}
\phi, & i=1,\dots,k,\\
0, & i=k+1,\dots,N.
\end{cases}
\end{equation}
The resulting $N$-photon state is thus expressed as
\begin{equation}
\ket{\psi_0}
=
\frac{1}{2^{N/2}}
\prod_{i=1}^{N}
\left(
\hat{a}_i^\dagger
+
\ee^{\ii\phi_i}\hat{b}_i^\dagger
\right)
\ket{0}.
\end{equation}
The QFT transforms the creation operators as
\begin{equation}
    \hat{\alpha}_{i}^{\dagger}
    \rightarrow
    \frac{1}{\sqrt{N}}
    \sum_{j=1}^{N}
    \ee^{\ii\frac{2\pi}{N}ij}\hat{\alpha}_{j}^{\dagger},
    \qquad \alpha=a,b .
\end{equation}
The resulting output state is
\begin{equation}
	\ket{\psi}
	=
	\frac{1}{(2N)^{N/2}}
	\prod_{i=1}^{N}
	\sum_{j=1}^{N}
	\left(
	\ee^{\ii\frac{2\pi}{N}ij}\hat{a}_{j}^{\dagger}
	+
	\ee^{\ii\frac{2\pi}{N}ij+\ii\phi_i}\hat{b}_{j}^{\dagger}
	\right)
	\ket{0}.
\end{equation}
It is useful to evaluate the commutator associated with a generic output photon $\alpha_p$, detected in the $p$th mode of $\alpha=a,b$, which is given by
\begin{equation}\label{eq}
	C_{\alpha,p;i,j}
	=
	\left[
	\hat{\alpha}_{p},
	\left(
	\ee^{\ii\frac{2\pi}{N}ij}\hat{a}_j^\dagger
	+
	\ee^{\ii\frac{2\pi}{N}ij+\ii\phi_i}\hat{b}_j^\dagger
	\right)
	\right]
	=
	\delta_{p,j}\ee^{\ii\frac{2\pi}{N}ij}
	\left(
	\delta_{\alpha,a}
	+
	\ee^{\ii\phi_i}\delta_{\alpha,b}
	\right)
	=
	\delta_{p,j}\ee^{\ii\frac{2\pi}{N}ip+\ii\delta_{\alpha,b}\phi_i},
	\quad \alpha=a,b .
	\end{equation}
Applying this commutator to the output state, we obtain
\begin{equation}
\begin{aligned}
	\hat{\alpha}_{p}\ket{\psi}
	&=
	\frac{1}{(2N)^{N/2}}
	\sum_{s_1=1}^{N}\sum_{j'=1}^{N}
	C_{\alpha,p;s_1,j'}
	\prod_{\substack{i=1\\i\neq s_1}}^{N}
	\sum_{j=1}^{N}
	\left(
	\ee^{\ii\frac{2\pi}{N}ij}\hat{a}_{j}^{\dagger}
	+
	\ee^{\ii\frac{2\pi}{N}ij+\ii\phi_i}\hat{b}_{j}^{\dagger}
	\right)
	\ket{0}
\\
	&=
	\frac{1}{(2N)^{N/2}}
	\sum_{s_1=1}^{N}
	\ee^{\ii\frac{2\pi}{N}s_1p+i\delta_{\alpha,b}\phi_{s_1}}
	\prod_{\substack{i=1\\i\neq s_1}}^{N}
	\sum_{j=1}^{N}
	\left(
	\ee^{\ii\frac{2\pi}{N}ij}\hat{a}_{j}^{\dagger}
	+
	\ee^{\ii\frac{2\pi}{N}ij+\ii\phi_i}\hat{b}_{j}^{\dagger}
	\right)
	\ket{0}.
\end{aligned}
\end{equation}
We can see from Eq.~\eqref{eq} that the features of each detected photon, characterized by the index $p$ and the mode $\alpha$, contribute independently to the phase of the commutator: the former acts as a "frequency" index of the quantum Fourier transform, while the latter serves as a "condition-checking" parameter that adds the phase $\phi_i$ only when the mode $\alpha=b$ is observed. On the other hand, a generic measurement outcome is defined by the number of photons detected in each output channel, represented by the sequence $\vec{n}=(n_{1,a},n_{1,b},\dots,n_{N,a},n_{N,b})$, with $\sum_{p=1}^N\sum_{\alpha=a,b} n_{p,\alpha}=N$. 
It is therefore convenient to construct, for a given sequence $\vec{n}$, two distinct objects: the ordered sequence of the indices $p$ of the modes in which each photon is observed,
\begin{equation}
	\vec{\gamma}=(\underbrace{1,\dots,1}_{n_{1,a}},\underbrace{1,\dots,1}_{n_{1,b}},\dots,\underbrace{N,\dots,N}_{n_{N,a}},\underbrace{N,\dots,N}_{n_{N,b}})=(\gamma_i)_{i=1,\dots,N},
\end{equation}
and the set $\mu$ of indices $t$ such that $\gamma_t$ corresponds to a photon observed in the mode $\alpha=b$. For $N=2$, we obtain
\begin{align}
    \vec{n}=(n_{1,a},n_{1,b},n_{2,a},n_{2,b})&\rightarrow\vec{\gamma}=(\gamma_1,\gamma_2),\ \mu=\{t|\gamma_t\text{ is mode b}\}:\notag\\
	\vec{n}=(2,0,0,0)&\rightarrow \vec{\gamma}=(1,1),\ \mu=\{\emptyset\}\notag\\
	\vec{n}=(0,2,0,0)&\rightarrow \vec{\gamma}=(1,1),\ \mu=\{1,2\}\notag\\
	\vec{n}=(0,0,2,0)&\rightarrow \vec{\gamma}=(2,2),\ \mu=\{\emptyset\}\notag\\
	\vec{n}=(0,0,0,2)&\rightarrow \vec{\gamma}=(2,2),\ \mu=\{1,2\}\notag\\
	\vec{n}=(1,1,0,0)&\rightarrow \vec{\gamma}=(1,1),\ \mu=\{2\}\notag\\
	\vec{n}=(1,0,1,0)&\rightarrow \vec{\gamma}=(1,2),\ \mu=\{\emptyset\}\notag\\
	\vec{n}=(1,0,0,1)&\rightarrow \vec{\gamma}=(1,2),\ \mu=\{2\}\notag\\
	\vec{n}=(0,1,1,0)&\rightarrow \vec{\gamma}=(1,2),\ \mu=\{1\}\notag\\
	\vec{n}=(0,1,0,1)&\rightarrow \vec{\gamma}=(1,2),\ \mu=\{1,2\}\notag\\
	\vec{n}=(0,0,1,1)&\rightarrow \vec{\gamma}=(2,2),\ \mu=\{2\}.
    \label{eq:Association}
\end{align}

Notice that, with the above definitions of $\vec{\gamma}$ and $\mu$, not all possible combinations are allowed. For example, although $\vec{\gamma}=(1,1)$ and $\mu={1}$ can be defined independently, this combination is excluded because the ordering of $\vec{\gamma}$ is chosen to avoid representing the same physical outcome more than once. Thus, the combination $\vec{\gamma}=(1,1)$ and $\mu={1}$ would correspond to the detection event $\vec{n}=(1,1,0,0)$, which is already represented by $\vec{\gamma}=(1,1)$ and $\mu={2}$. The probability $P_{\vec{n}}\equiv P_{\vec{\gamma},\mu}$ of observing the outcome $\vec{n}$ is therefore
\begin{multline}
	P_{\vec{n}} = \frac{1}{\prod_{\substack{{p=1}\\{\alpha=a,b}}}^Nn_{p,\alpha}!}|\bra{0}\prod_{\substack{{p=1}\\{\alpha=a,b}}}^N \hat{\alpha}_p^{n_{p,\alpha}}\ket{\psi}|^2=
    \frac{1}{(2N)^{N}\prod_{\substack{{p=1}\\{\alpha=a,b}}}^Nn_{p,\alpha}!}\left\vert\sum_{s_1=1}^N\ee^{\ii \frac{2\pi}{N}s_1 \gamma_1+\ii\delta_{\alpha_1
    ,b}\phi_{s_1}}\sum_{\substack{{s_2=1}\\{s_2\neq s_1}}}^N\ee^{\ii \frac{2\pi}{N}s_2 \gamma_2+\ii\delta_{\alpha_2
    ,b}\phi_{s_2}}\cdots\right\vert^2=\\
    =\mathcal{N}_{\vec{n}}\left\vert\sum_{\sigma\in S(N)} \ee^{\ii \frac{2\pi}{N}\vec{\sigma}\cdot\vec{\gamma}+\ii\sum\limits_{t\in\mu}\phi_{\sigma_t}}\right\vert^2=\mathcal{N}_n\left\vert\sum_{\sigma\in S(N)} \ee^{\ii \frac{2\pi}{N}\vec{\sigma}\cdot\vec{\gamma}+\ii\phi\mathcal{M}_{\mu,\sigma}}\right\vert^2
\end{multline}
where $\mathcal{N}_{\vec{n}}= ((2N)^{N}\prod_{\substack{{p=1}\\{\alpha=a,b}}}^Nn_{p,\alpha}!)^{-1}$, $S(N)$ denotes the symmetric group of permutations of the indices $1,2,\dots,N$, and 
$\vec{\sigma}\cdot\vec{\gamma}=\sum_{i=1}^{N}\sigma_i\gamma_i$. 
For a given permutation $\sigma$, $\mathcal{M}_{\mu,\sigma}$ counts the number of indices satisfying $1\leq\sigma_t\leq k$ for $t\in\mu$, and can be written as
\begin{equation}
	\mathcal{M}_{\mu,\sigma}
	=\sum_{t\in\mu}\chi_k(\sigma_t)
	=\sum_{t=1}^{N}\xi_\mu(t)\chi_k(\sigma_t),
\end{equation}
where
\begin{equation}
\chi_k(r)=
\begin{cases}
1, & 1\leq r\leq k,\\
0, & k<r\leq N,
\end{cases}
\qquad
\xi_\mu(t)=
\begin{cases}
1, & t\in\mu,\\
0, & t\notin\mu.
\end{cases}
\end{equation}
Finally, in the small-phase regime $\phi\simeq0$, we expand the probability as
\begin{equation}
	P_{\vec{n}}\simeq\mathcal{N}_{\vec{n}}
	\sum_{\sigma,\sigma'\in S(N)}
	\ee^{\ii\frac{2\pi}{N}(\vec{\sigma}-\vec{\sigma}')\cdot\vec{\gamma}}
	\left[
	1+\ii\phi(\mathcal{M}_{\mu,\sigma}-\mathcal{M}_{\mu,\sigma'})
	-\frac{1}{2}\phi^2(\mathcal{M}_{\mu,\sigma}-\mathcal{M}_{\mu,\sigma'})^2
	\right].
\end{equation}
\subsection{Vanishing First-Order Contribution and Relevant Outcomes}

We first verify that $\partial_\phi P_{\vec{n}}\big|_{\phi=0}=0$ for all $\vec{n}$. Consequently, only the outcomes satisfying $P_{\vec{n}}=0$ contribute non-vanishingly to the CFI. Indeed, we have
\begin{equation}
\begin{aligned}
	\partial_\phi P_{\vec{n}}
	&\simeq
	i\,\mathcal{N}_{\vec{n}}
	\sum_{\sigma,\sigma'\in S(N)}
	\ee^{\ii\frac{2\pi}{N}(\vec{\sigma}-\vec{\sigma}')\cdot\vec{\gamma}}
	\left(\mathcal{M}_{\mu,\sigma}
	-\mathcal{M}_{\mu,\sigma'}\right)
\\[2mm]
	&=
	i\,\mathcal{N}_{\vec{n}}
	\sum_{\sigma,\sigma'\in S(N)}
	\ee^{\ii\frac{2\pi}{N}(\vec{\sigma}-\vec{\sigma}')\cdot\vec{\gamma}}
	\left(\mathcal{M}_{\mu,\sigma+s}
	-\mathcal{M}_{\mu,\sigma'+s}\right),
\end{aligned}
\end{equation}

In the second equality, we perform the change of variables
$\sigma\rightarrow\sigma+s$ and $\sigma'\rightarrow\sigma'+s$, where the addition is understood element-wise modulo $N$. Since this transformation is a relabeling of the permutations, it leaves the value of $\partial_\phi P_{\vec{n}}$ unchanged.
\begin{equation}
\partial_\phi P_{\vec{n}}\simeq
\ii\,\mathcal{N}_{\vec{n}}\frac{1}{N}
\sum_{\sigma,\sigma'\in S(N)}
\ee^{\ii\frac{2\pi}{N}(\vec{\sigma}-\vec{\sigma}')\cdot\vec{\gamma}}
\sum_{s=0}^{N-1}
(\mathcal{M}_{\mu,\sigma+s}-\mathcal{M}_{\mu,\sigma'+s}).
\end{equation}
We further observe that
\begin{equation}
\sum_{s=0}^{N-1}\mathcal{M}_{\mu,\sigma+s}
=
\sum_{t=1}^{N}\xi_\mu(t)
\sum_{s=0}^{N-1}\chi_k(\sigma_t+s)
=
k\sum_{t=1}^{N}\xi_\mu(t),
\end{equation}
where the last equality follows from the fact that, as $s$ spans all values from $0$ to $N-1$, the quantity $\sigma_t+s$ covers all indices from $1$ to $N$ exactly once, with only $k$ of them belonging to the interval $[1,k]$. Therefore,
\begin{equation}
\begin{aligned}
\partial_\phi P_{\vec{n}}
&\simeq
\ii\,\mathcal{N}_{\vec{n}}\frac{1}{N}
\sum_{\sigma,\sigma'\in S(N)}
\ee^{\ii\frac{2\pi}{N}(\vec{\sigma}-\vec{\sigma}')\cdot\vec{\gamma}}
\left(
k\sum_{t=1}^{N}\xi_\mu(t)
-
k\sum_{t=1}^{N}\xi_\mu(t)
\right)
\\
&=0 .
\end{aligned}
\end{equation}
Thus, to evaluate the CFI, it is sufficient to consider only the outcomes $\vec{n}$ for which $P_{\vec{n}}=0$. For these events, the probability near $\phi\simeq0$ is given by
\begin{equation}
P_{\vec{n}}\simeq
-\mathcal{N}_{\vec{n}}\frac{\phi^2}{2}
\sum_{\sigma,\sigma'\in S(N)}
\ee^{\ii\frac{2\pi}{N}(\vec{\sigma}-\vec{\sigma}')\cdot\vec{\gamma}}
(\mathcal{M}_{\mu,\sigma}-\mathcal{M}_{\mu,\sigma'})^2 .
\label{eq:Prob0}
\end{equation}

\subsection{Identifying the Outcomes Contributing to the CFI}
One can verify that outcomes $\vec{n}^{\,*}$ satisfying
\begin{equation}
	\sum_{i=1}^N \gamma_i^* \neq 0 \quad (\mathrm{mod}\ N),
    \label{eq:ConditionGamma}
\end{equation}
also satisfy $P_{\vec{n}^*}\big\vert_{\phi=0}=0$, or equivalently
\begin{equation}
	\sum_{\sigma\in S(N)} \ee^{\ii \frac{2\pi}{N}\vec{\sigma}\cdot\vec{\gamma}^*}=0.
    \label{eq:PropGammaS}
\end{equation}
Indeed, if $\vec{\sigma}+\vec{1}$ denotes the permutation obtained by adding the vector $\vec{1}=(1,\dots,1)$ element-wise modulo $N$, the invariance of the sum over all permutations under this transformation gives
\begin{equation}
	\sum_{\sigma\in S(N)}
	\ee^{\ii \frac{2\pi}{N}\vec{\sigma}\cdot\vec{\gamma}^{\,*}}
	=
	\sum_{\sigma\in S(N)}
	\ee^{\ii \frac{2\pi}{N}(\vec{\sigma}+\vec{1})\cdot\vec{\gamma}^{\,*}}
	=
	\ee^{\ii \frac{2\pi}{N}\sum_i\gamma_i^*}
	\sum_{\sigma\in S(N)}
	\ee^{\ii \frac{2\pi}{N}\vec{\sigma}\cdot\vec{\gamma}^{\,*}}.
\end{equation}
which implies that, whenever $\vec{\gamma}^{\,*}$ satisfies Eq.~\eqref{eq:ConditionGamma}, Eq.~\eqref{eq:PropGammaS} also holds. Since Eq.~\eqref{eq:ConditionGamma} is only a sufficient condition, there may exist outcomes $\vec{\bar{n}}$ for which $P_{\vec{\bar{n}}}=0$ while $\sum_{i=1}^{N}\bar{\gamma}_i=0$. In the following, we show that only outcomes of the type $\vec{n}^{\,*}$ contribute to the CFI.
\subsection{Deriving the CFI from the Contributing Outcomes}

Here, we evaluate the contribution to the CFI from the outcomes $\vec{n}^{\,*}$ satisfying $\sum_i\gamma_i^{\,*}\neq 0 \pmod N$. Denoting this contribution by $F^*$, and using Eq.~\eqref{eq:Prob0}, which applies since $P_{\vec{n}^{\,*}}=0$, we obtain
\begin{equation}
\begin{aligned}
	F^*
	&=
	\sum_{\vec{n}^{\,*}}
	\frac{\left(\partial_\phi P_{\vec{n}^{\,*}}\right)^2}
	{P_{\vec{n}^{\,*}}}
\\[2mm]
	&=
	-2
	\sum_{\vec{n}^{\,*}}
	\mathcal{N}_{\vec{n}^{\,*}}
	\sum_{\sigma,\sigma'\in S(N)}
	\ee^{\ii \frac{2\pi}{N}(\vec{\sigma}-\vec{\sigma}')\cdot\vec{\gamma}^{\,*}}
	\left(\mathcal{M}_{\mu,\sigma}
	-\mathcal{M}_{\mu,\sigma'}\right)^2
\\[2mm]
	&=
	4
	\sum_{\vec{n}^{\,*}}
	\mathcal{N}_{\vec{n}^{\,*}}
	\sum_{\sigma,\sigma'\in S(N)}
	\ee^{\ii \frac{2\pi}{N}(\vec{\sigma}-\vec{\sigma}')\cdot\vec{\gamma}^{\,*}}
	\mathcal{M}_{\mu,\sigma}
	\mathcal{M}_{\mu,\sigma'},
\end{aligned}
\end{equation}

where, in the last step, we have used Eq.~\eqref{eq:PropGammaS}. To perform the sum over $\vec{n}^{\,*}$, it is convenient to identify all possible combinations of $\vec{\gamma}^{\,*}$ and $\mu$ associated with a physical outcome while counting each outcome only once. To this end, we allow $\vec{\gamma}^{\,*}$ to take any value in $(1,\dots,N)^{\otimes N}$ and $\mu$ to be any subset of $\{1,\dots,N\}$, correcting for the resulting overcounting by dividing each contribution by its multiplicity. Specifically, each outcome
$\vec{n}=(n_{1,a},n_{1,b},\dots,n_{N,a},n_{N,b})$
is counted
$N!/\prod_{i=1}^{N}(n_{i,a}!n_{i,b}!)$
times, corresponding to the number of distinct permutations of the $N$ detected photons. For example, when $N=2$, the only outcomes requiring this correction are those in which the two photons occupy different output modes, as these are counted twice. This can be verified directly from the list of associations in Eq.~\eqref{eq:Association}.
\begin{align}
    \vec{n}=(1,1,0,0)&\leftarrow [\vec{\gamma}=(1,1),\ \mu=\{2\};\vec{\gamma}=(1,1),\ \mu=\{1\}]\notag\\
	\vec{n}=(1,0,1,0)&\leftarrow [\vec{\gamma}=(1,2),\ \mu=\{\emptyset\};\vec{\gamma}=(2,1),\ \mu=\{\emptyset\}]\notag\\
	\vec{n}=(1,0,0,1)&\leftarrow [\vec{\gamma}=(1,2),\ \mu=\{2\};\vec{\gamma}=(2,1),\ \mu=\{1\}]\notag\\
	\vec{n}=(0,1,1,0)&\leftarrow [\vec{\gamma}=(1,2),\ \mu=\{1\};\vec{\gamma}=(2,1),\ \mu=\{2\}]\notag\\
	\vec{n}=(0,1,0,1)&\leftarrow [\vec{\gamma}=(1,2),\ \mu=\{1,2\};\vec{\gamma}=(2,1),\ \mu=\{1,2\}]\notag\\
	\vec{n}=(0,0,1,1)&\leftarrow [\vec{\gamma}=(2,2),\ \mu=\{2\};\vec{\gamma}=(2,2),\ \mu=\{1\}].
\end{align}
It is now straightforward to perform the sum over all outcomes $\vec{n}^{\,*}$, yielding
\begin{equation}
	F^*
    =
	4\mathcal{N}'
	\sum_{\sigma,\sigma'\in S(N)}
	\left(
	\sum_{\substack{\vec{\gamma}^{\,*}\in\{1,\dots,N\}^{\otimes N}\\
	\sum_i\gamma_i^{\,*}\neq0}}
	\ee^{\ii \frac{2\pi}{N}(\vec{\sigma}-\vec{\sigma}')\cdot\vec{\gamma}^{\,*}}
	\right)
	\left(
	\sum_{\mu\in2^{\{1,\dots,N\}}}
	\mathcal{M}_{\mu,\sigma}\mathcal{M}_{\mu,\sigma'}
	\right),
	\label{eq:boundFI}
\end{equation}
where
\begin{equation}
        \mathcal{N}'= \mathcal{N}_n \prod_{i=1}^N\frac{n_{i,a}!n_{i,b}!}{N!}=
    \frac{1}{(2N)^{N}N!}
\end{equation}
is independent of the particular outcome, and $2^{S}$ denotes the power set of $S$, so that $\mu$ runs over all the subsets of $\{1,\dots,N\}$.
From now on, we will shorten the notation of the two summations by writing $\sum_{\gamma^*}$ and $\sum_{\mu}$. We notice that
\begin{equation}
	\sum_\mu\mathcal{M}_{\mu,\sigma}\mathcal{M}_{\mu,\sigma'}= 
    \sum_{t,t'=1}^{N}\sum_\mu\xi_\mu(t)\xi_\mu(t')\chi_k(\sigma_t)\chi_k(\sigma'_{t'})
\end{equation}
and, in particular, $\sum_\mu\xi_\mu(t)\xi_\mu(t')$ counts the number of sets $\mu$ for which both indices $t$ and $t'$ belong to $\mu$:
\begin{equation}
	\sum_\mu\xi_\mu(t)\xi_\mu(t')=
    \begin{cases}
        \sum_\mu\xi_\mu(t)^2=\sum_\mu\xi_\mu(t)=2^{N-1} & \mathrm{for}\ t=t'\\
        2^{N-2} & \mathrm{for}\ t\neq t'
    \end{cases}
    =2^{N-2}(1+\delta_{t,t'}).
\end{equation}
We can substitute this result in
\begin{equation}
	\sum_\mu\mathcal{M}_{\mu,\sigma}\mathcal{M}_{\mu,\sigma'}=
    2^{N-2}\left[
    \sum_{t=1}^N\chi_k(\sigma_t)\sum_{t'=1}^N\chi_k(\sigma'_{t'})
    +\sum_{t=1}^N\chi_k(\sigma_t)\chi_k(\sigma'_t)
    \right]
\end{equation}
and, since
\begin{equation}
\sum_{t=1}^N\chi_k(\sigma_t)=k,
\end{equation}
we obtain
\begin{equation}
	\sum_\mu\mathcal{M}_{\mu,\sigma}\mathcal{M}_{\mu,\sigma'}=
    2^{N-2}\left(k^2+\sum_{t=1}^N\chi_k(\sigma_t)\chi_k(\sigma'_t)\right).
\end{equation}
Therefore,
\begin{equation}
\begin{aligned}
	F^*
	&= 
    \mathcal{N}'2^{N}\sum_{\sigma,\sigma'\in S(N)}
    \left(\sum_{\gamma^*}\ee^{\ii \frac{2\pi}{N}(\vec{\sigma}-\vec{\sigma}')\cdot\vec{\gamma}^*}\right)
    \left(k^2+\sum_{t=1}^N\chi_k(\sigma_t)\chi_k(\sigma'_t)\right)\\
	&=
    \mathcal{N}'2^{N}\sum_{\sigma,\sigma'\in S(N)}
    \left(\sum_{\gamma^*}\ee^{\ii \frac{2\pi}{N}(\vec{\sigma}-\vec{\sigma}')\cdot\vec{\gamma}^*}\right)
    \sum_{t=1}^N\chi_k(\sigma_t)\chi_k(\sigma'_t),
\end{aligned}
\end{equation}
where, in the last step, we used the property of $\gamma^*$ given in Eq.~\eqref{eq:PropGammaS}. We now exploit the orthogonality relation
\begin{equation}
	\sum_{\gamma}\ee^{\ii \frac{2\pi}{N}(\vec{\sigma}-\vec{\sigma}')\cdot\vec{\gamma}_n}
    = N^N\delta_{\sigma,\sigma'},
\end{equation}
which, however, requires the sum to be taken over all $N$-tuples $\gamma$, and not only over the subset $\gamma^*$ satisfying Eq.~\eqref{eq:ConditionGamma}. Let $\tilde{\gamma}$ denote the $N$-tuples such that $\sum_i\tilde{\gamma}_i=0$ mod $N$, i.e., all the remaining tuples not included in $\gamma^*$. Then,

\begin{equation}
\begin{aligned}
	F^*
	&=
    \mathcal{N}'2^{N}\sum_{\sigma,\sigma'\in S(N)}
    \left(
    N^N\delta_{\sigma,\sigma'}
    -\sum_{\tilde{\gamma}}
    \ee^{\ii \frac{2\pi}{N}(\vec{\sigma}-\vec{\sigma}')\cdot\vec{\tilde{\gamma}}}
    \right)
    \sum_{t=1}^N\chi_k(\sigma_t)\chi_k(\sigma'_t)\\
	&=
    k-
    \mathcal{N}'2^N\sum_{\sigma,\sigma'\in S(N)}
    \left(
    \sum_{\tilde{\gamma}}
    \ee^{\ii \frac{2\pi}{N}(\vec{\sigma}-\vec{\sigma}')\cdot\vec{\tilde{\gamma}}}
    \right)
    \sum_{t=1}^N\chi_k(\sigma_t)\chi_k(\sigma'_t).
\end{aligned}
\end{equation}
Here, $\sum_{\tilde{\gamma}}$ is a shorthand notation for the sum over all $\tilde{\gamma}$ satisfying $\sum_i\tilde{\gamma}_i=0$ mod $N$. We now rewrite the constrained sum over $\tilde{\gamma}$ as an unconstrained sum by exploiting the identity
\begin{equation}
	\frac{1}{N}\sum_{s=0}^{N-1} 
    \ee^{\ii \frac{2\pi}{N} s\sum_{i}\gamma_i}
    =
    \begin{cases}
        1 & \mathrm{if}\ \sum_i\gamma_i = 0\ \mathrm{mod}\ N,\\
        0 & \mathrm{otherwise}.
    \end{cases}
\end{equation}
This allows us to write
\begin{equation}
\begin{aligned}
	\sum_{\tilde{\gamma}}
    \ee^{\ii \frac{2\pi}{N}(\vec{\sigma}-\vec{\sigma}')\cdot\vec{\tilde{\gamma}}}
    &=
    \sum_{\gamma}
    \ee^{\ii \frac{2\pi}{N}(\vec{\sigma}-\vec{\sigma}')\cdot\vec{\gamma}}
    \frac{1}{N}\sum_{s=0}^{N-1}
    \ee^{\ii \frac{2\pi}{N}s\sum_i\gamma_i}\\
    &=
    \frac{1}{N}\sum_{\gamma}\sum_{s=0}^{N-1}
    \ee^{\ii \frac{2\pi}{N}
    (\vec{\sigma}-\vec{\sigma}'+s\vec{1})\cdot\vec{\gamma}}\\
    &=
    N^{N-1}\sum_{s=0}^{N-1}\delta_{\sigma+s,\sigma}.
\end{aligned}
\end{equation}
Therefore,
\begin{equation}
\begin{aligned}
	F^*
	&=
    k-\frac{1}{N!N}
    \sum_{s=0}^{N-1}\sum_{\sigma\in S(N)}
    \sum_{t=1}^N
    \chi_k(\sigma_t)\chi_k((\sigma+s)_t).
\end{aligned}
\end{equation}
Here, $\sigma+s$ denotes the element-wise addition of $s$ to $\sigma$ modulo $N$. Finally, since the sum over $s$ runs from $0$ to $N-1$, the index $\sigma_t+s$ also spans all values from $0$ to $N-1$. Hence,
\begin{equation}
\begin{aligned}
	F^*
	&=
    k-\frac{1}{N!N}
    \sum_{\sigma\in S(N)}
    \sum_{t=1}^N
    \chi_k(\sigma_t)
    \sum_{s=0}^{N-1}\chi_k((\sigma+s)_t)\\
	&=
    k-\frac{k^2}{N}\\
	&=
    \frac{k(N-k)}{N}.
\end{aligned}
\end{equation}

\subsection{Vanishing Contribution of Zero-Probability Outcomes}
We are now left with the task of proving that any outcome $\vec{\bar{n}}$ such that 
$P_{\vec{\bar{n}}}=0$ and $\sum_i\bar{\gamma}_i=0$ mod($N$) gives no contribution to the Fisher information, i.e., $F_{\bar{n}}=0$ for all $\bar{n}$. We can still write
\begin{equation}
\begin{aligned}
	F_{\bar{n}}
	&=
    4\mathcal{N}'\sum_{\sigma,\sigma'\in S(N)}
    \ee^{\ii \frac{2\pi}{N}(\vec{\sigma}-\vec{\sigma}')\cdot\vec{\bar{\gamma}}}
    \mathcal{M}_{\mu,\sigma}\mathcal{M}_{\mu,\sigma'}\\
	&=
    4\mathcal{N}'\sum_{\sigma,\sigma'\in S(N)}
    \ee^{\ii \frac{2\pi}{N}(\vec{\sigma}-\vec{\sigma}')\cdot\vec{\bar{\gamma}}}
    \sum_{t,t'=1}^N
    \chi_k(\sigma_t)\chi_k(\sigma'_{t'})
    \xi_\mu(t)\xi_\mu(t').
\end{aligned}
\end{equation}
We now perform the change of variables $\sigma\rightarrow\sigma+s$ and 
$\sigma'\rightarrow\sigma'+s'$. Using the property
$\sum_i\bar{\gamma}_i=0$ mod($N$), we have
\[
(s-s')\vec{1}\cdot\vec{\bar{\gamma}}
=
(s-s')\sum_i\bar{\gamma}_i
=
0\ \mathrm{mod}(N),
\]
and therefore, the exponential factor remains unchanged. We obtain
\begin{equation}
	F_{\bar{n}}
	=
    4\mathcal{N}'\sum_{\sigma,\sigma'\in S(N)}
    \ee^{\ii \frac{2\pi}{N}(\vec{\sigma}-\vec{\sigma}')\cdot\vec{\bar{\gamma}}}
    \sum_{t,t'=1}^N
    \chi_k(\sigma_t+s)\chi_k(\sigma'_{t'}+s')
    \xi_\mu(t)\xi_\mu(t').
\end{equation}
Since this relation holds for every $s$ and $s'$, we can replace the expression with its average over both variables:
\begin{equation}
	F_{\bar{n}}
	=
    4\mathcal{N}'\frac{1}{N^2}
    \sum_{\sigma,\sigma'\in S(N)}
    \ee^{\ii \frac{2\pi}{N}(\vec{\sigma}-\vec{\sigma}')\cdot\vec{\bar{\gamma}}}
    \sum_{t,t'=1}^N
    \sum_{s=0}^{N-1}\chi_k(\sigma_t+s)
    \sum_{s'=0}^{N-1}\chi_k(\sigma'_{t'}+s')
    \xi_\mu(t)\xi_\mu(t').
\end{equation}
Using again the relation
\[
\sum_{s=0}^{N-1}\chi_k(\sigma_t+s)=k,
\]
which is independent of $\sigma$, we obtain
\begin{equation}
\begin{aligned}
	F_{\bar{n}}
	&=
    4k^2\mathcal{N}'\frac{1}{N^2}
    \underbrace{
    \sum_{\sigma,\sigma'\in S(N)}
    \ee^{\ii \frac{2\pi}{N}(\vec{\sigma}-\vec{\sigma}')\cdot\vec{\bar{\gamma}}}
    }_{0}
    \sum_{t,t'=1}^N\xi_\mu(t)\xi_\mu(t')=0
\end{aligned}
\end{equation}
The vanishing of the underbraced term follows from the fact that 
$P_{\bar{n}}=0$ for these outcomes (and therefore Eq.~\eqref{eq:PropGammaS} must also hold for $\vec{\bar{n}}$). Therefore, we have proven that
\begin{equation}
	F=F^*=\frac{k(N-k)}{N}\equiv\frac{kl}{N}.
\end{equation}

\section{Geometric Origin of the QFI Eigenstructure}\label{zl}
We consider the normalized state
\begin{equation}
|\psi\rangle
=
\frac{1}{\sqrt{M}}
\left(
|1\rangle
+
\sum_{k=2}^{M} \ee^{\ii\phi_k}|k\rangle
\right),
\end{equation}
and introduce a single-parameter phase deformation
\(\phi_k \rightarrow \phi_k + \theta s_k\),
where \(s=(s_2,\ldots,s_M)\) is a normalized real vector. This defines the family
\(|\psi(\theta)\rangle = e^{\ii\theta \hat G_s}|\psi\rangle\),
with generator $\hat G_s = \sum_{k=2}^{M} s_k |k\rangle\langle k|$. The corresponding tangent vector is
\begin{equation}\nonumber
|\partial_\theta \psi\rangle
=
\frac{i}{\sqrt{M}}
\sum_{k=2}^{M} s_k \ee^{\ii\phi_k}|k\rangle,
\end{equation}
with
\(
\langle \partial_\theta \psi|\partial_\theta \psi\rangle = \frac{1}{M}
\)
and overlap
\(
\langle \psi|\partial_\theta \psi\rangle
=
\frac{i}{M}\sum_{k=2}^{M}s_k.
\)
The quantum Fisher information (QFI) then takes the form
\begin{equation}
F_Q
=
\frac{4}{M}
-
\frac{4}{M^2}
\left(\sum_{k=2}^{M} s_k\right)^2,
\end{equation}
which explicitly separates the contributions that are parallel and orthogonal to the state. For relative phase modes satisfying \(\sum_{k=2}^{M}s_k=0\), the tangent vector is orthogonal to the state, yielding
\begin{equation}\nonumber
F_{\rm rel}=\frac{4}{M}.
\end{equation}
For the symmetric mode \(s_k = 1/\sqrt{M-1}\), the deformation has a component parallel to the state, resulting in
\begin{equation}\nonumber
F_{\rm sym}=\frac{4}{M^2}.
\end{equation}
Geometrically, the QFI measures only the component of the state variation orthogonal to the state vector. Relative phase deformations fully explore the projective Hilbert space and therefore achieve higher sensitivity, whereas symmetric deformations are partially aligned with the state vector, leading to a reduced QFI scaling.
\section{Fisher Information for Local Homodyne Detection}\label{mx}
The single-photon state can be described by introducing an effective mode
\[
c^\dagger = \frac{1}{\sqrt{2}}\left(a^\dagger + e^{\ii\phi} b^\dagger\right),
\]
such that the state corresponds to a single-photon excitation in mode \(c\). The associated Wigner function is given by
\[
W(x_c,p_c) = \frac{1}{\pi} e^{-x_c^2 - p_c^2}\left(2x_c^2 + 2p_c^2 - 1\right).
\]
To express it in terms of the original quadratures $(x_a,p_a,x_b,p_b)$, we use the fact that the transformation defining $c^\dagger$ is part of a unitary transformation
\begin{equation}\nonumber
    \begin{pmatrix}
        c^\dagger\\
        d^\dagger
    \end{pmatrix}
    =
    \frac{1}{\sqrt{2}}
    \begin{pmatrix}
        1 & \ee^{\ii\phi}\\
        1 & -\ee^{\ii\phi}
    \end{pmatrix}
    \begin{pmatrix}
        a^\dagger\\
        b^\dagger
    \end{pmatrix}
    =
    U^*
    \begin{pmatrix}
        a^\dagger\\
        b^\dagger
    \end{pmatrix},
\end{equation}
which induces a symplectic rotation $z' = R z$ in phase space
\begin{equation}\nonumber
    R=
    \begin{pmatrix}
        \mathrm{Re}(U)&-\mathrm{Im}(U)\\
        \mathrm{Im}(U)&\mathrm{Re}(U)
    \end{pmatrix}
    =
    \frac{1}{\sqrt{2}}
    \begin{pmatrix}
        1&\cos\phi&0&\sin\phi\\
        1&-\cos\phi&0&-\sin\phi\\
        0&-\sin\phi&1&\cos\phi\\
        0&\sin\phi&1&-\cos\phi
    \end{pmatrix}.
\end{equation}
The corresponding quadrature transformations are given by
\[
x_c = \frac{1}{\sqrt{2}}\left(x_a + x_b\cos\phi + p_b\sin\phi\right), \qquad
p_c = \frac{1}{\sqrt{2}}\left(p_a - x_b\sin\phi + p_b\cos\phi\right).
\]
Applying the inverse quadrature transformation to the two-mode extension of the Wigner function gives
\[
W(x_a,p_a,x_b,p_b) = \frac{1}{\pi^2} e^{-R^2}
\left(R^2 - 1 + 2\cos\phi (x_a x_b + p_a p_b) + 2\sin\phi (x_a p_b - p_a x_b)\right),
\]
where $R^2 = x_a^2 + p_a^2 + x_b^2 + p_b^2$. The marginal distribution associated with the joint homodyne measurement of $x_a$ and $p_b$ is obtained by integrating over $x_b$ and $p_a$
\[
P(x_a,p_b) = \int dx_b\, dp_a \, W(x_a,p_a,x_b,p_b)
= \frac{1}{\pi} e^{-x_a^2 - p_b^2} \left(x_a^2 + p_b^2 + 2 x_a p_b \sin\phi \right).
\]
This distribution reproduces the relevant moments, e.g.
\[
\mathbb{E}[x_a^{2n+1} p_b^{2m+1}] 
= 2\sin\phi \, \frac{(2n+1)!! (2m+1)!!}{2^{n+m+2}},
\quad
\mathbb{E}[x_a^{2n} p_b^{2m}] 
= \frac{(2n-1)!! (2m-1)!!}{2^{n+m}} (n+m+1).
\]
The CFI is then given by
\[
F_H(\phi) = \frac{1}{\pi} \int dx_a\, dp_b \, e^{-x_a^2 - p_b^2}
\frac{4 x_a^2 p_b^2 \cos^2\phi}{x_a^2 + p_b^2 + 2 x_a p_b \sin\phi}.
\]
Maximizing over $\phi$ gives $F_H(0)=1/2$, which coincides with the single-photon limit $F(N=2,k=1)$. This establishes the ultimate phase sensitivity attainable using local homodyne detection. To analyze the effects of loss, we introduce the density operator
\begin{equation}
\rho=(1-\epsilon)\ket{0}\bra{0}+\frac{\epsilon}{2}(a^\dagger+\ee^{\ii\phi}b^\dagger)\ket{0}\bra{0}(a+\ee^{-\ii\phi}b),
\label{eq:LossyState}
\end{equation}
which represents an incoherent mixture of the vacuum and a single-photon superposition. Consequently, the homodyne probability distribution is
\begin{equation}
P_\epsilon(x_a,p_b)=\frac{1}{\pi}e^{-x_a^2-p_b^2}\left[1-\epsilon+\epsilon\left(x_a^2+p_b^2+2x_a p_b\sin\phi\right)\right].
\end{equation}
The corresponding CFI is
\begin{equation}
F_{H,\epsilon}(\phi)=\frac{1}{\pi}\int dx_a\,dp_b\, e^{-x_a^2-p_b^2}
\frac{4\epsilon^2 x_a^2 p_b^2 \cos^2\phi}{1-\epsilon+\epsilon(x_a^2+p_b^2+2x_a p_b\sin\phi)}.
\end{equation}
In the weak-signal limit $\epsilon\ll1$, this expression simplifies to
\begin{equation}
F_{H,\epsilon}(\phi)\xrightarrow{\epsilon\to 0}
\frac{1}{\pi}\int dx_a\,dp_b\, e^{-x_a^2-p_b^2}
\,4\epsilon^2 x_a^2 p_b^2 \cos^2\phi
=\epsilon^2\cos^2\phi,
\end{equation}
recovering the quadratic scaling with the transmission parameter.
\section{Phase Estimation with Three-Mode Homodyne Detection}\label{pz}
We consider the three-mode single-photon W state
\begin{equation}\nonumber
|\psi\rangle
=
\zeta^\dagger|0\rangle,
\qquad
\zeta^\dagger
=
\frac{1}{\sqrt{3}}
\left(
a^\dagger
+
e^{\ii\phi_2} b^\dagger
+
e^{\ii\phi_3} c^\dagger
\right),
\end{equation}
where mode \(a\) defines the phase reference \(\phi_1=0\). The corresponding Wigner function associated with the excitation operator \(\zeta\) is
\begin{equation}\nonumber
W(x_\zeta,p_\zeta)
=
\frac{1}{\pi}
e^{-x_\zeta^2-p_\zeta^2}
\left(
2x_\zeta^2+2p_\zeta^2-1
\right).
\end{equation}
The collective quadratures are related to the local quadratures through
\begin{align*}\nonumber
x_\zeta
&=
\frac{1}{\sqrt{3}}
\Big(
x_a
+x_b\cos\phi_2+p_b\sin\phi_2
+x_c\cos\phi_3+p_c\sin\phi_3
\Big),\\
p_\zeta
&=
\frac{1}{\sqrt{3}}
\Big(
p_a
-x_b\sin\phi_2+p_b\cos\phi_2
-x_c\sin\phi_3+p_c\cos\phi_3
\Big).
\end{align*}
To account for arbitrary local-oscillator phases
\(\{\theta_1,\theta_2,\theta_3\}\), we define the effective phase differences
\begin{equation}\nonumber
\bar\phi_k=\phi_k-\theta_k,
\qquad
k\in\{1,2,3\},
\end{equation}
with \(\phi_1=0\). In terms of these effective phases, the rotated collective quadratures take the compact form
\begin{align*}\nonumber
x_\zeta
&=
\frac{1}{\sqrt{3}}
\sum_{s=1}^3
\left(
x_s\cos\bar\phi_s
+
p_s\sin\bar\phi_s
\right),\\
p_\zeta
&=
\frac{1}{\sqrt{3}}
\sum_{s=1}^3
\left(
-p_s\sin\bar\phi_s
+
p_s\cos\bar\phi_s
\right).
\end{align*}
Upon substituting the rotated collective quadratures into the single-photon Wigner function, the full three-mode Wigner quasiprobability distribution is obtained,
\begin{equation}\nonumber
	W(\boldsymbol{z}) = \frac{1}{\pi^3} e^{-R^2}(\frac{2}{3}R^2-1+\frac{4}{3}(\sum_{s>t=1}^3 \cos(\bar{\phi}_s-\bar{\phi}_t)(x_s x_t + p_s p_t)+\sin(\bar{\phi}_s-\bar{\phi}_t)(x_t p_s - x_s p_t))
\end{equation}
where $R^2=\sum_{s=1}^3(x_s^2+p_s^2)$. The interference terms are proportional to
\(\cos(\bar\phi_s-\bar\phi_t)\)
and
\(\sin(\bar\phi_s-\bar\phi_t)\)
contain all phase correlations among the three modes. The dependence on pairwise phase differences is a direct manifestation of the coherent superposition underlying the three-mode $W$ state. Integrating the Wigner function over all momentum quadratures then yields the joint probability distribution for local homodyne measurements,
\begin{multline}\nonumber
	P(x_a,x_b,x_c) = \int d p_a d p_b d p_c\ W(z)\\
    = \frac{2}{3\pi^{3/2}} e^{-x_a^2-x_b^2-x_c^2}\left(x_a^2+x_b^2+x_c^2+2x_a x_b\cos(\phi_2+\theta_1-\theta_2)+2x_a x_c\cos(\phi_3+\theta_1-\theta_3)+2x_b x_c\cos(\phi_2-\phi_3-\theta_2+\theta_3)\right)
\end{multline}
The cross terms \(x_i x_j\) originate from quantum interference between different excitation pathways, with their amplitudes determined by the corresponding cosine factors. To account for photon loss, we introduce a transmission parameter \(\epsilon\ll1\), resulting in
\begin{equation}\nonumber
	P_\epsilon=\frac{2}{3\pi^{3/2}} e^{-x_a^2-x_b^2-x_c^2}\left(1-\epsilon+\epsilon(x_a^2+x_b^2+x_c^2+2x_a x_b\cos(\phi_2+\theta_1-\theta_2)+2x_a x_c\cos(\phi_3+\theta_1-\theta_3)+2x_b x_c\cos(\phi_2-\phi_3-\theta_2+\theta_3))\right)
\end{equation}
The CFI matrix thus reads
\begin{equation}\nonumber
(F_\epsilon)_{kl}
=
\int dx_a\,dx_b\,dx_c\;
\frac{\partial_{\phi_k}P_\epsilon\,\partial_{\phi_l}P_\epsilon}{P_\epsilon},
\qquad k,l=2,3.
\end{equation}
In the high photon-loss regime \(\epsilon \ll 1\), this expression reduces to
\begin{equation}\nonumber
F_\epsilon
=
\epsilon^{2}
\int dx_a\,dx_b\,dx_c\;
\frac{8}{3\pi^{3/2}}
e^{-x_a^{2}-x_b^{2}-x_c^{2}}
\,\mathcal{M},
\end{equation}
where
\begin{equation}\nonumber
\mathcal{M}
=
\begin{pmatrix}
A^{2} & AB \\
AB & B^{2}
\end{pmatrix},
\end{equation}
with
\begin{align*}
A &= x_a x_b \sin(\bar{\phi}_1-\bar{\phi}_2)
- x_b x_c \sin(\bar{\phi}_2-\bar{\phi}_3), \\
B &= x_a x_c \sin(\bar{\phi}_1-\bar{\phi}_3)
+ x_b x_c \sin(\bar{\phi}_2-\bar{\phi}_3).
\end{align*}
Performing the Gaussian integration gives the leading-order result

\begin{equation}
    F_\epsilon(\phi)
    =
    \frac{2}{3}\epsilon^2
    \begin{pmatrix}
        \sin^2(\bar{\phi}_1-\bar{\phi}_2)
        +\sin^2(\bar{\phi}_2-\bar{\phi}_3)
        &
        -\sin^2(\bar{\phi}_2-\bar{\phi}_3)
        \\[4pt]
        -\sin^2(\bar{\phi}_2-\bar{\phi}_3)
        &
        \sin^2(\bar{\phi}_1-\bar{\phi}_3)
        +\sin^2(\bar{\phi}_2-\bar{\phi}_3)
    \end{pmatrix}
    =
    \begin{pmatrix}
        F_{\phi_1} & F_{\phi_1\phi_2}\\
        F_{\phi_1\phi_2} & F_{\phi_2}
    \end{pmatrix},
\end{equation}
In the symmetric configuration,  $\phi_2=\phi_3\equiv\phi$, the two-parameter estimation reduces to a single-parameter problem. The associated CFI is obtained by combining the contributions from the original phase parameters
\begin{equation*}
    F_\epsilon(\phi)
    =
    \frac{1}{P_\epsilon}
    \bigl(\partial_\phi P_\epsilon\bigr)^2
    =
    \frac{1}{P_\epsilon}
    \bigl(\partial_{\phi_1}P_\epsilon+\partial_{\phi_2}P_\epsilon\bigr)^2.
\end{equation*}
Using the matrix elements, one obtains
\begin{equation*}
    F_\epsilon(\phi)
    =
    F_{\phi_1}
    +
    F_{\phi_2}
    +
    2F_{\phi_1\phi_2}
    =
    \frac{2}{3}\epsilon^2
    \left[
        \sin^2(\bar{\phi}_1-\bar{\phi}_2)
        +
        \sin^2(\bar{\phi}_1-\bar{\phi}_3)
    \right].
\end{equation*}
Optimal sensitivity is achieved when
\[
\phi_2-\theta_2+\theta_1
=
\phi_3-\theta_3+\theta_1
=
\frac{\pi}{2},
\]
leading to
\begin{equation}\nonumber
F_\epsilon(\phi)
=
\frac{4}{3}\epsilon^{2}.
\end{equation}
Thus, homodyne detection achieves optimal sensitivity in the phase quadrature, while the scaling $F_\epsilon(\phi) \propto \epsilon^2$, highlights the severe degradation of phase information caused by photon loss.
\end{document}